\documentclass[12pt, a4paper]{article}

\usepackage[height=25cm,width=16cm]{geometry}
\usepackage{amsfonts}
\usepackage{amsmath}
\usepackage{amssymb}
\usepackage{dcolumn}
\usepackage{bm,here}
\usepackage{subfig}
\usepackage{comment}
\usepackage{ifpdf}
\usepackage{slashed}
\usepackage{colortbl}
\usepackage{color}
\usepackage[normalem]{ulem}
\usepackage[mathscr]{eucal}
\usepackage{cite}
\usepackage[utopia]{mathdesign}
\usepackage[sort&compress, numbers, merge]{natbib}

\usepackage[utf8]{inputenc}

\usepackage{cancel}

\ifpdf
  \usepackage{graphicx}     
  \usepackage[bookmarksopen,colorlinks=true,linkcolor=bblue,citecolor=ppink,urlcolor=ppink,linktocpage=false]{hyperref}
\else     
  \usepackage[dvipdfmx]{graphicx}     
  \usepackage[dvipdfmx,bookmarksopen,colorlinks=true,linkcolor=bblue,citecolor=ppink,urlcolor=ppink,linktocpage=false]{hyperref}
\fi

\usepackage{multicol}
\definecolor{red}{rgb}{1,0,0}
\definecolor{ppink}{rgb}{0.921545,0.440586,0.687243}
\definecolor{bblue}{rgb}{0.400000,0.400000,1.000000}

\newcommand{\vev}[1]{ \left< {#1} \right> }

\newcommand{\com}[1]{\left[ {#1} \right]}

\begin{document}

\begin{titlepage}

\begin{flushright}
IPMU16-0039 \\
LAPTH-013/16
\end{flushright}

\begin{center}

\vskip 2.5cm
{\Large \bf WIMP Dark Matter in a Well-Tempered Regime}\\[1em]
{\large \bf A case study on Singlet-Doublets Fermionic WIMP}
\vskip 1.5cm

{\large
Shankha Banerjee$^{(a)}$,
Shigeki Matsumoto$^{(b)}$,\\[.5em]
Kyohei Mukaida$^{(b)}$
and Yue-Lin Sming Tsai$^{(b)}$
}

\vskip 1.0cm
\begin{tabular}{l}
$^{(a)}$ {\sl LAPTH, Univ. de Savoie, CNRS, B.P.110, F-74941 Annecy-le-Vieux, France}\\[.3em]
$^{(b)}$ {\sl Kavli IPMU (WPI), UTIAS, University of Tokyo, Kashiwa, 277-8583, Japan}
\end{tabular}

\vskip 2.0cm
\begin{abstract}
\noindent
Serious searches for the weakly interacting massive particle (WIMP) have now begun. In this context, the most important questions that need to be addressed are: {\it To what extent can we constrain the WIMP models in the future?} and {\it What will then be the remaining unexplored regions in the WIMP parameter space for each of these models?}  
In our quest to answer these questions, we classify WIMP in terms of quantum number
and study each case adopting minimality as a guiding principle.
As a first step,
we study one of the simple cases of the minimal composition in the well-tempered fermionic WIMP regime, namely the singlet-doublets WIMP model. We consider all available constraints from direct and indirect searches and also the predicted constraints coming from the near future and the future experiments. We thus obtain the current status, the near future prospects and the future prospects of this model in all its generality. We find that in the future,  this model will be constrained almost solely by the future direct dark matter detection experiments (as compared to the weaker indirect and collider constraints) and the cosmological (relic density) constraints and will hence be gradually pushed to the corner of the coannihilation region, if no WIMP signal is detected. Future lepton colliders will then be useful in exploring this region not constrained by any other experiments.
\end{abstract}

\end{center}

\end{titlepage}

\tableofcontents
\thispagestyle{empty}
\newpage
\setcounter{page}{1}

\section{Introduction}
\label{sec: intro}

There is no doubt about the existence of dark matter in our universe thanks to astrophysical and cosmological observations ranging from galactic to cosmological scales. The nature of dark matter, however, still remains as one of the biggest mysteries of particle physics in spite of tremendous efforts to ascertain its nature since several decades. What we currently know is that the dark matter in our universe has an abundance of $\Omega_c h^2 = 0.1198 (15)$\,\cite{Ade:2015xua} at present and that it interacts gravitationally. The dark matter is expected to be some new particle which is electrically neutral and colorless. Among several candidates proposed so far, the Weakly Interacting Massive Particle (WIMP) is one of the most attractive candidates because of its simplicity and predictability. The thermal freeze-out mechanism of WIMP, which is nowadays called the WIMP miracle, naturally explains the abundance observed today when its mass is around the electroweak scale. This may open up a way to understand the origin of the electroweak symmetry breaking (EWSB). Moreover, we are hopeful to detect the WIMP, because some of the predicted interactions between the WIMP and the standard model (SM) particles provide a strong driving force of WIMP detections at the high energy colliders, underground laboratories (direct dark matter detections) and in cosmological/astrophysical observations (indirect dark matter detections).

We have, on the other hand, not obtained any conclusive signatures of the WIMP so far, and it pushes the celebrated WIMP 
scenarios gradually to the corner. This fact, however, also means that the era of serious WIMP searches has begun, so that currently the most important questions are {\it To what extent can we constrain the WIMP models in the future?} and {\it What will then be the remaining unexplored regions in the WIMP parameter space for each of these models?} If the simplicity and predictability of the WIMP are getting lost, such theories will become less attractive. If not, we have to consider what kinds of experiments are required to cover the unconstrained parameter region. Several big experiments, such as multi-tons scale direct dark matter detections\,\cite{Malling:2011va, Kish:2014daa} and future lepton colliders\,\cite{Behnke:2013xla, CEPC-SPPCStudyGroup:2015csa, Gomez-Ceballos:2013zzn}, are being proposed, and it is worth figuring out what roles these experiments can play in this direction. The goal of this paper is to try and answer certain aspects of the aforementioned questions by performing a comprehensive analysis of the singlet-doublets WIMP model~\cite{ArkaniHamed:2006mb}.

We take a standard strategy to study such models based on the WIMP's quantum number. Once the spin of the WIMP is fixed, the WIMP field can always be written as a linear combination of colorless representations of the SM gauge group\footnote{A decomposition formula like Eq.~\ref{eq1} is always possible. Of course, if certain additional symmetries exist, one can extend this formula by including the additional representations. The only case which one must tread carefully is when the WIMP is vector-like. In this scenario, there will always be some additional gauge symmetry, and we should include the explicit form in our formula from the very beginning.}, \textit{viz.} SU(2)$_L \times$ U(1)$_Y$, which must involve the electrically neutral components:
\begin{align}
\label{eq1}
	\text{WIMP}(x) = \sum_i\,z_i\,[\chi_i(x)]_\text{N.\,C.},
\end{align}
with `N.\,C.' indicating an electrically neutral component of the representation $\chi_i(x)$. The coefficient $z_i$ follows the sum rule $\sum_i |z_i|^2 = 1$ because of the canonical normalization of the field, and $|z_i|^2$ measures how much the representation $\chi_i(x)$ is involved as a component of the WIMP field. The most straightforward way for studying such WIMP scenarios is to write down an effective Lagrangian of the WIMP field involving all interactions which can be responsible for the WIMP miracle phenomenon, though it is not practical because we have to introduce infinitely many representations and interactions. Instead, we try to cover the theory space by an infinite number of small patches based on the quantum number of the WIMP and construct an appropriate effective Lagrangian in each patch. However, there are still infinitely many possibilities of the effective Lagrangian in each patch; we can always construct the Lagrangian  with any desired complexity. We therefore take simplicity as a guiding principle to write down the Lagrangian in each patch.

The WIMP models can be divided into two categories: one in which the WIMP is described almost by a gauge eigenstate of the SU(2)$_L$ interaction, namely one of the $|z_i|^2$s is close to unity. The other is the one in which the WIMP is not described by a single SU(2)$_L$ gauge eigenstate. It indicates that none of the $|z_i|^2$s are close to unity. This is nowadays referred to as the well-tempered WIMP\,\cite{ArkaniHamed:2006mb}. The former category is further divided into many patches: when a SU(2)$_L$ singlet component dominates, the so called singlet WIMP is realized\,\cite{Silveira:1985rk, McDonald:1993ex, Jungman:1995df, Servant:2002aq, Kim:2008pp, Kakizaki:2006dz, Ellwanger:2009dp, Kanemura:2010sh, Djouadi:2011aa, Belanger:2010yx, Cline:2013gha, Matsumoto:2014rxa, Beniwal:2015sdl}. The bino-like neutralino in the minimal supersymmetric standard model (MSSM) and the Kaluza-Klein photon in the universal extra-dimension model (UED) are good examples. When the SU(2)$_L$ doublet components dominate,\footnote{In order to describe the fermionic doublet WIMP, we introduce a pair of SU(2)$_L$ doublet (Weyl) fermions whose hypercharges are opposite with each other, for it makes the theory free from anomaly.} the doublet WIMP is obtained\,\cite{Mizuta:1992qp, Ma:2006km, Barbieri:2006dq, Nagata:2014wma}. Examples of such a WIMP are the Higgsino-like neutralino in the MSSM and the inert Higgs doublet in some two Higgs doublet models (2HDMs). We have the triplet WIMP when a SU(2)$_L$ triplet component dominates\,\cite{Gherghetta:1999sw, Moroi:1999zb, Hisano:2006nn}, with a famous example being the wino-like neutralino in the MSSM. A WIMP which is described by a higher multiplet is nowadays called the minimal dark matter\,\cite{Cirelli:2005uq, Cirelli:2007xd}.

WIMP in the latter category, namely the well-tempered WIMP, is more complicated to be classified. A systematic method to decompose it into patches can be based on the number of representations that are participating in for describing the well-tempered WIMP. For example, it follows $|z_i|^2 + |z_j|^2 \simeq 1$ when this number is two with both $|z_i|^2$ and $|z_j|^2$ being not close to unity, $|z_i|^2 + |z_j|^2 + |z_k|^2 \simeq 1$ when the number is three, and so on. The first example, say the minimal composition in this paper, is usually adequate enough to grasp the typical nature of the well-tempered WIMP. Moreover, it is actually realized in most of the parameter regions of realistic new physics models, even though we should not deny interesting possibilities that WIMPs described by many representations can potentially show some new features that are not seen in the minimal composition. Here, it is worth pointing out how coannihilation regions are incorporated in this framework. This framework automatically involves coannihilations between components inside a SU(2)$_L$ multiplet and between particles which are mixed with each other in the well-tempered WIMP. On the other hand, other coannihilations, such as the one between WIMP and a particle which resides in a different SU(2)$_L$ multiplet and is not mixed with the WIMP, should be considered in each individual patch, even though such a coannihilation region usually does not appear in the minimal setup of the effective Lagrangian in each patch.

In this paper, we focus on a fermionic WIMP in the well-tempered regime.  In the next section (Sec.\,\ref{sec: framework}) the well-tempered WIMP in general, and show that some non-trivial conditions are required in the minimal composition setup. After that, the simplest case of the minimal composition setup, which is called the singlet-doublets WIMP model in this paper, is presented.\footnote{Other examples of the minimal composition setup utilizing higher SU(2)$_L$ multiplets, such as the doublets-triplet WIMP model\,\cite{Dedes:2014hga} and the triplet-quadruplets WIMP model\,\cite{Tait:2016qbg}, have also been studied.} We discuss the so-called Higgs- and $Z$-boson-blind spot parameter regions of the model which are recently gaining attention because they evade severe direct dark matter detection constraints\,\cite{Cheung:2012qy}. The WIMP model has indeed been studied in the literature\,\cite{Mahbubani:2005pt, D'Eramo:2007ga, Cohen:2011ec, Cheung:2013dua, Calibbi:2015nha, Freitas:2015hsa, Fedderke:2015txa, NewOne}, but a complete analysis is still lacking. We want to show the current status of the model in its general parameter space, how it will change in the near future and what will be the parameter space remaining in the future if no WIMP signal is detected by then. Current experimental constraints and their expected future improvements are summarized in Sec.\,\ref{sec: constraints}, which involves those from dark matter relic density, direct dark matter detections, indirect dark matter detections and also those from high energy colliders. We also discuss the detailed physics behind these constraints. Following these, in Sec.\,\ref{sec: results}, we perform a numerical analysis to scan the parameter space of the WIMP model and figure out its (would-be) status at present, near future and future, respectively, by imposing the experimental constraints discussed in the previous section. Sec.\,\ref{sec: summary} is devoted to a summary of our findings and discusses the kind of experiments and observations required to explore the yet unconstrained parameter regions in the future. For the sake of completeness, we also put a detailed discussion on the blind spot regions in Appendix\,\ref{app: blind spots}, a supplementary explanation about a collider study of the well-tempered WIMP at the Large Hadron Collider (LHC) with high luminosity in Appendix\,\ref{app: lhc14}, and several figures in Appendix\,\ref{app: figures} which are obtained by scanning the model parameter space numerically and supplement the figures in the main text.

\section{Well-Tempered WIMP}
\label{sec: framework}

As was mentioned in the introduction, the WIMP is in general described by a linear combination of colorless representations of the SM gauge group $\chi_i$; WIMP $= \sum_i z_i\,[\chi_i]_{\rm N.\,C.}$ with $[\chi_i]_{\rm N.\,C.}$ being an electrically neutral component of $\chi_i$. We focus on a fermionic WIMP in the well-tempered regime, where all of the $|z_i|^2$ are not very close to one. In order to make the following discussion concrete, we define the regime as follows:
\begin{eqnarray}
R_i \equiv |z_i|^2 \leq 0.95, \qquad {\rm for~~all}~~i.
\label{eq: well-tempered condition}
\end{eqnarray}
Though the WIMP can potentially be composed of many representations $\chi_i$, it is usually described by a minimal composition in most of the realistic scenarios. We thus take the number of participating representations 
describing the WIMP to be as small as possible.

Then, the representations for the well-tempered WIMP must follow several criteria. First, we must introduce at least two representations whose weak isospins differ by one half, because a large enough mixing between different representations calls for a Yukawa interaction with the SM Higgs field whose weak isospin is 1/2. Second, because one of the two representations, which has a half integer weak isospin, also carries a non-zero hypercharge so that it contains an electrically neutral component, its conjugate field must also be introduced in order to make the theory anomaly free. Finally, the other representation, which has an integer weak isospin, must have a Majorana mass term. Otherwise the WIMP becomes a Dirac particle giving a large scattering cross-section off a nucleus through a $Z$-boson exchange, which has already been ruled out by recent direct dark matter detection experiments. In summary, the hypercharge of the representation having an integer weak isospin must be zero, while those of the other representations having a half integer weak isospin are 1/2 and -1/2 due to the aforementioned reason. 

We introduce the following three representations as a minimal composition: a representation having an integer weak isospin and no hypercharge, and two representations (which are conjugates of each other) having a half integer weak isospin and hypercharges of $\pm$1/2. The minimal composition is then expressed as follows:
\begin{align}
	{\bf (2n - 1)}_0, \, {\bf (2n)}_{1/2}, \, {\bf (2n)}_{-1/2}
	\qquad {\rm or} \qquad
	{\bf (2n + 1)}_0, \, {\bf (2n)}_{1/2}, \, {\bf (2n)}_{-1/2},
\end{align}
with $n$ being an integer equal to or larger than one. The numbers in parentheses stand for the quantum number of the SU(2)$_L$ interaction, while the subscript is the hypercharge $Y$. Some examples of field contents in the minimal scenario are shown in Tab.~\ref{tab: minimal composition}. We will consider the simplest one as a case study of a WIMP in the well-tempered regime, where it involves a SU(2)$_L$ singlet Weyl fermion field with $Y = 0$ ($S$) and two SU(2)$_L$ doublet Weyl fermion fields with $Y = \pm 1/2$ ($D_1$ and $D_2$). We will clarify how severely the present dark matter search experiments put constraints on the model parameter space of the WIMP, and discuss some prospects to detect the WIMP in the (near) future.

\renewcommand{\arraystretch}{1.5}
\begin{table}[t]
\begin{center}
	\begin{tabular}{lcccr} \hline
	&Weyl Fermion & SU(2)$_L$ & SU(3)$_C$ & U(1)$_Y$ \\
	\hline
	\hline
	\rowcolor[cmyk]{0,0,0.1,0}
	Singlet-Doublets& $S$ & ${\bf 1}$ & ${\bf 1}$& $0$ \\
	\rowcolor[cmyk]{0,0,0.1,0}
	& $D_1$ & ${\bf 2}$ & ${\bf 1}$ & $1/2$ \\
	\rowcolor[cmyk]{0,0,0.1,0}
	& $D_2$ & ${\bf 2}$ & ${\bf 1}$ & $- 1/2$ \\
	\hline
	\rowcolor[cmyk]{0.1,0,0,0}
	Doublets-Triplet & $D_1$ & ${\bf 2}$ & ${\bf 1}$ & $1/2$ \\
	\rowcolor[cmyk]{0.1,0,0,0}& $D_2$ & ${\bf 2}$ & ${\bf 1}$ & $- 1/2$ \\
	\rowcolor[cmyk]{0.1,0,0,0}& $T$ & ${\bf 3}$ & ${\bf 1}$& $0$ \\
	\hline
	Triplet-Quadruplets& $T$ & ${\bf 3}$ & ${\bf 1}$& $0$ \\
	& $Q_1$ & ${\bf 4}$ & ${\bf 1}$ & $1/2$ \\
	& $Q_2$ & ${\bf 4}$ & ${\bf 1}$ & $- 1/2$ \\
	\hline
	\end{tabular}
	\caption{\it \small
	Some examples of field contents in the minimal composition of a well-tempered WIMP.}
	\label{tab: minimal composition}
\end{center}
\end{table}
\renewcommand{\arraystretch}{1}

\subsection{Singlet-Doublets WIMP Model}
\label{sec: sd}

The singlet-doublets model for the well-tempered WIMP, which involves three Weyl fermion fields, \textit{viz.} $S$, $D_1$ and $D_2$, is described by the following Lagrangian:
\begin{align}
	{\cal L}_\text{SD} 
	&= {\cal L}_\text{kin}
	- \com{ \frac{1}{2} M_S S S 
	+ M_D D_1 \cdot D_2
	+ y_1 S D_1 \cdot \tilde H
	+ y_2 S D_2 \cdot H 
	+ \text{H.c.}
	},
	\label{eq: sd model}
\end{align}
with ${\cal L}_\text{kin}$ being the kinetic term of the three new fermion fields, whose form will be explicitly mentioned later in the text. The Lagrangian involves all renormalizable interactions of the three fermion fields by taking the $Z_2$ odd property of the fields into account.\footnote{A $Z_2$ symmetry is implicitly imposed in the Lagrangian\,(\ref{eq: sd model}) in order to guarantee the stability of the WIMP. The three new particles are odd under this symmetry, while all of the SM particles are even under it.} The dot ($\cdot$) indicates the contraction of the SU(2)$_L$ indices via the anti-symmetric tensor $\epsilon_{ij}$. The SM Higgs field is denoted by $H$ with its quantum number ${\bf 2}_{1/2}$, while $\tilde H \equiv \epsilon H^\dag$ is its conjugate with the quantum number ${\bf 2}_{-1/2}$. We suppress all possible higher dimensional interactions that might come from integrating out some other new heavy particles in a fundamental theory, assuming these particles are enough heavier than the fermions in the above Lagrangian. The above model can thus be regarded as an effective theory describing physics around and below the mass scale of the three fermions.

We have four complex parameters $y_1$, $y_2$, $M_S$ and $M_D$. Three of their phases are rotated away by redefining the three Weyl fermion fields, and one phase remains as a physical one associated with the invariant $\phi \equiv \arg[ M_S M_D y_1^\ast y_2^\ast ]$. When the phase is neither $0$ nor $\pi$, the interactions of the model violate CP symmetry and contribute to the electric dipole moment of electron through Barr-Zee type diagrams\,\cite{Mahbubani:2005pt, D'Eramo:2007ga, Giudice:2005rz}, which have already been limited by recent experiments\,\cite{Baron:2013eja} and will be constrained even more severely in the near future\,\cite{Sakemi:2011zz, Kawall:2011zz, Kara:2012ay}. We thus take $\phi = 0$ or $\pi$ in our analysis to avoid such constraints. The model is hence characterized by four real parameters with one sign: $y_1$, $y_2$, $M_S$, $M_D$ and $\text{sign} \com{M_S M_D y_1 y_2 } = \pm 1$. We take the following convention to describe our results. First, $M_S$, $y_1$, $y_2 \geqslant 0$, while $M_D$ being both positive and negative to take the sign into account. Next, we parametrize the two Yukawa couplings as $y_1 = y \cos \theta$ and $y_2 = y \sin \theta$ with $y \geqslant 0$ and $0 \leqslant \theta \leqslant \pi/2$. The range of the angle $\theta$ is further restricted to be $\pi/4 \leqslant \theta \leqslant \pi/2$ because of the symmetry which leaves physics described by the model unchanged: $(y_1, y_2) \leftrightarrow (y_2 , y_1)$. Concerning the symmetry, it is worth noting that we can always rename the doublet fields as $(D_1, D_2) \leftrightarrow (\bar D_2, \bar D_1)$. In summary, we have four real parameters in the model: {\boldmath $M_S \geqslant 0$}, {\boldmath $M_D$}, {\boldmath $y \geqslant 0$} and {\boldmath $\pi/4 \leqslant \theta \leqslant \pi/2$} ({\boldmath $\tan \theta \geqslant 1$} or {\boldmath $0 \leqslant \cot \theta \leqslant 1$}).

It is here instructive to compare our model with the MSSM\footnote{
We must note here that the singlet-doublets model has overlap only with a particular limit of MSSM, where only binos and higgsinos lighter compared to the other SUSY breaking parameters. 
}. In the MSSM, our singlet field corresponds to the Bino, while the doublet fields are the up- and down-type Higgsinos. Moreover, we have a correspondence between our model parameters and those of the MSSM: $M_S \leftrightarrow M_1$, $M_D \leftrightarrow \mu$, $\tan \theta \leftrightarrow \tan \beta$ and $y \leftrightarrow g'/\sqrt{2}$. Here, $M_1$, $\mu$, $g'$ and $\tan \beta$ are the supersymmetry breaking Bino mass, the supersymmetry invariant Higgsino mass, the U(1)$_Y$ gauge coupling and the ratio between the vacuum expectation values of two Higgs doubles introduced in the MSSM. The Difference thus appears at the strength of the Yukawa coupling; it is described by the gauge coupling $g'$ in the MSSM, while it is taken to be a free parameter in our model.

\subsection{Interactions in the Singlet-Doublets WIMP Model}

We have so far discussed the setup of the singlet-doublets WIMP model. In what follows, we consider the mass spectra and interactions predicted by the model.

\subsubsection{Mass spectra}

The Higgs field acquires the vacuum expectation value after the electroweak symmetry breaking as $\vev{H} = (0, v)^T/\sqrt{2}$ with $v \sim 246$\,GeV. As a result, the neutral components of the singlet and doublet fields are mixed with each other. Using the notation $D_1 = (D_1^+, D_1^0)^T$ and $D_2 = (D_2^0, D_2^-)^T$, the mass terms of the neutral components are
\begin{align}
	{\cal L}_N = - \frac{1}{2} (S, D_1^0, D_2^0)\,M_N
	\begin{pmatrix} S \\ D_1^0 \\ D_2^0 \end{pmatrix} + \text{H.c.},
	\qquad
	M_N \equiv \begin{pmatrix} M_S & -y_1 v/\sqrt{2} & y_2 v/\sqrt{2} \\
	-y_1 v/\sqrt{2} & 0 & -M_D \\ y_2 v/\sqrt{2} & -M_D & 0 \end{pmatrix},
	\label{eq: mass matrix}
\end{align}
while those of the charged components are simply given by {\boldmath ${\cal L}_C = - M_D D_1^+ D_2^- + \text{\bf H.c.}$} Since $M_N$ is a real symmetric matrix, it can be diagonalized by an orthogonal matrix $O_N$. In order to make all mass eigenvalues positive, we consider an unitary matrix $U_N \equiv O_N\,\Phi$ instead of $O_N$, where $\Phi = {\rm diag} (\eta_1, \eta_2, \eta_3)$ with each component taking a value 1 or $i$ according to the eigenvalue of $O_N^T M_N O_N$. Then, the mass eigenvalues and eigenstates are
\begin{align}
	U_N^T M_N U_N = {\rm diag} (M_{N1}, M_{N2}, M_{N3}),
	\qquad
	\tilde{N} \equiv U_N^\dag (S, D_1^0, D_2^0)^T,
\end{align}
where we adopt the convention $M_{N3} \geqslant M_{N2} \geqslant M_{N1} \geqslant 0$. Here, it is useful to introduce the parameters $R_S$ and $R_D$ in order to quantify how the singlet and doublet components are mixed with each other in the lightest $Z_2$ odd state described by the field $\tilde{N}_1$:
\begin{align}
	R_S \equiv |(U_N)_{11}|^2, \qquad
	R_D \equiv |(U_N)_{21}|^2 + |(U_N)_{31}|^2,
	\label{eq: mixing parameters}
\end{align}
where $R_S + R_D = 1$ is guaranteed by the unitarity of $U_N$, or in other words, the normalization of the lightest $Z_2$ odd field. As was already stated in Eq.\,(\ref{eq: well-tempered condition}), we consider the parameter region satisfying $R_S, R_D \leqslant 0.95$ in order to focus on the WIMP in the well-tempered regime.

\subsubsection{Interactions}

For the sake of convenience, instead of the two component spinor notation used so far, we use the following four component notation for the interactions of the model (\ref{eq: sd model}): 
\begin{align}
	N_i \equiv (\{\tilde{N}_i^\dag\}_\alpha,\,\tilde{N}_i^{\dot \alpha})^T, \qquad
	C \equiv (\{D_2^{-\,\dag}\}_\alpha,\,\{D^+_1\}^{\dot \alpha})^T.
\end{align}
Using the four component spinors, their kinetic terms are simply given by canonical forms {\boldmath $(1/2)\,\bar{N}_i\,(i\slashed{\partial} - M_{N_i})\,N_i$} and {\boldmath $\bar{C}\,(i\slashed{\partial} - M_D)\,C$}. On the other hand, the interactions of the model after the electroweak symmetry breaking are summarized as follows:
\begin{align}
	{\cal L}_{\rm int} =&
	-h\,\bar{N}_i\,[y_{ij} P_L + y_{ij}^\ast P_R]\,N_j
	-\bar{N}_i\,\slashed{Z}\,[g_{ij} P_L - g_{ij}^\ast P_R]\,N_j
	+e\,\bar{C}\,\slashed{A}\,C 
	+\frac{g}{2c_W}\,(2c_W^2 - 1)\,\bar{C}\,\slashed{Z}\,C \nonumber \\
	&
	+\frac{g}{\sqrt{2}}\,\bar{C}\,\slashed{W}^\dag\,[(U_N)_{2i} P_L - (U_N^\ast)_{3i} P_R]\,N_i
	+\frac{g}{\sqrt{2}}\,\bar{N}_i\,\slashed{W}\,[(U_N^\dag)_{i2} P_L - (U_N^T)_{i3} P_R]\,C, 	
\end{align}
where $h$, $\slashed{Z} = Z_\mu \gamma^\mu$, $\slashed{A} = A_\mu \gamma^\mu$ and $\slashed{W} = W_\mu \gamma^\mu$ are the Higgs boson, photon, $Z$ and $W$ boson fields, respectively, with $P_{L/R} = (1/2)\,(1 \mp \gamma_5)$ being the chirality projection operator, while $e$, $g$ and $c_W = \cos \theta_W$ are the electromagnetic coupling, the weak gauge coupling and the Weinberg angle, respectively. The couplings $y_{ij}$ and $g_{ij}$ are defined as
\begin{align}
	y_{ij} &\equiv \frac{1}{\sqrt{2}} [-(U_N^T)_{i1} y_1 (U_N)_{2j} + (U_N^T)_{i1} y_2 (U_N)_{3j}],
	\label{eq: def y11} \\
	g_{ij} &\equiv \frac{g}{4 c_W} [(U_N^\dag)_{i2} (U_N)_{2j} - (U_N^\dag)_{i3} (U_N)_{3j}].
	\label{eq: def g11}
\end{align}

\subsubsection{Blind spots}
\label{subsubsec: blind spots}

It is instructive to discuss the couplings $y_{11}$ and $g_{11}$ in some details, because they play important roles in the WIMP phenomenology developed in the following sections.

We first consider the coupling $y_{11}$, which represents the interaction strength between the WIMP and the Higgs boson, and leads to the spin-independent scattering of the WIMP off a nucleus. After some calculations (see Appendix\,\ref{app: higgs blind spot}), its explicit form reads
\begin{align}
	y_{11} = y_{11}^\ast =
	- \frac{y^2\,v\,(\eta_1^2 M_D \sin 2\theta + M_{N1})}
	{2M_D^2 + y^2 v^2 + 4 \eta_1^2 M_S M_{N1} - 6 M_{N1}^2}.
	\label{eq: y11}
\end{align}
Since we assumed the CP invariance in the model, the Yukawa coupling $y_{ii}$ becomes real ($y_{ii} = y_{ii}^\ast$), so that all the neutral particles $N_i$s have their own scalar interactions ($h\,\bar{N}_i N_i$). On the other hand they do not have the pseudoscalar ones ($i h\,\bar{N}_i \gamma_5 N_i$).

This coupling $y_{11}$ has already been severely constrained by several present direct dark matter detection experiments, as will be shown in the next section. The parameter region with $y_{11} \simeq 0$ is thus preferred, which is nowadays known to be {\bf the Higgs blind spot region}\,\cite{Cheung:2012qy}. The condition $y_{11} = 0$ holds when
\begin{align}
	M_{N1} = - \eta_1^2 M_D \sin 2 \theta.
	\label{eq: higgs blind spot}
\end{align}
Moreover, it turns out that the WIMP mass becomes $M_{N1} = M_S$ and $\eta_1 = 1$ when this condition holds (see again Appendix\,\ref{app: higgs blind spot}). As a result, the blind spot region appears when the mass parameter of the charged particle $C$ is negative, namely $M_D < 0$. The other neutral particles $N_2$ and $N_3$ have masses $M_{N2} = M_{N3} = (M_D^2 + y^2 v^2 /2)^{1/2}$. The mixing parameter $R_S$ in Eq.\,(\ref{eq: mixing parameters}) is given by {\boldmath $R_S = (M_D^2 - M_{N1}^2)/(M_D^2 - M_{N1}^2 + y^2 v^2/2)$} in this region, so that the condition in Eq.\,(\ref{eq: higgs blind spot}) can be accommodated with the one in Eq.\,(\ref{eq: well-tempered condition}).

Next we consider the coupling $g_{11}$, which gives the axial gauge coupling of the WIMP to the $Z$-boson and leads to the spin-dependent scattering between the WIMP and a nucleus. Its explicit from is obtained by using the same method for $y_{11}$ (see Appendix\,\ref{app: z blind spot}):
\begin{align}
	g_{11} = g_{11}^\ast = 
	- \frac{y^2\,v\,m_Z\,\cos 2\theta}
	{2\,(2M_D^2 + y^2 v^2 + 4 \eta_1^2 M_S M_{N1} - 6 M_{N1}^2)},
	\label{eq: g11}
\end{align}
which again gives a real coupling ($g_{ii} = g_{ii}^\ast$), so that all the neutral particles $N_i$s have their own axial vector current interactions ($\bar{N}_i \slashed{Z} \gamma_5 N_i$), while they never have the vector current ones ($\bar{N}_i \slashed{Z} N_i$), as expected from the Majorana nature of the neutral particles $N_i$s.

This coupling is being gradually limited by recent direct dark matter detection experiments
and it will be more constrained in the near future if no signals are detected. Thus, the region with $g_{11} \sim 0$, which is called {\bf the $Z$-boson blind spot region}\,\cite{Cheung:2012qy}, will be preferred. According to the explicit form in Eq.\,(\ref{eq: g11}), the coupling $g_{11}$ is suppressed when $\theta \sim \pi/4$. Eventually, the regions satisfying both the blind spot conditions,
 Eqs.\,\eqref{eq: higgs blind spot} and $\theta \sim \pi/4$, will be preferred if no signals are detected.
As a result, all of the particles $N_1$, $N_2$, $N_3$ and $C$ tend to be degenerate. It is worth noticing here that a very degenerate spectrum leads to $R_S < 0.05$ as can be seen in the $R_S$ formula in the previous paragraph, and it contradicts with the condition for the well-tempered WIMP.

\section{Experimental Constraints}
\label{sec: constraints}

In this section, we discuss the existing experimental constraints used to constrain the model parameter region of the singlet-doublets WIMP model. Moreover, we also discuss some expected constraints which are likely to be obtained in the (near) future in dark matter search experiments, and figure out which part of the regions allowed by the existing constraints will be explored there. We employ the profile-likelihood method\,\cite{Rolke:2004mj} to search for the region with high probability, in which various experimental constraints are incorporated in the form of the likelihood function $L$ with their statistical and systematical uncertainties. The likelihood function that we adopt is constructed by four components:
\begin{eqnarray}
	L[M_S, M_D, y, \cot \theta] =
	L_{\rm CS}[M_S, \cdots] \times
	L_{\rm DD}[M_S, \cdots] \times
	L_{\rm CL}[M_S, \cdots] \times
	L_{\rm ID}[M_S, \cdots],
\end{eqnarray}
where these component likelihood functions $L_{\rm CS}$, $L_{\rm DD}$, $L_{\rm CL}$ and $L_{\rm ID}$ are constructed based on experimental results obtained from dark matter cosmology, direct dark matter detections, collider experiments and indirect dark matter detections, respectively. We evaluate the likelihood function, $L[M_S, M_D, y, \cot \theta]$, numerically using the so-called MultiNest sampling algorithm\,\cite{Feroz:2008xx}. In what follows, we will discuss all of the component functions in some details together with the physics behind them.

\subsection{Dark Matter Relic Density}
\label{subsec: CS}

We adopt the following likelihood function $L_{\rm CS}$, which is taken as a Gaussian:
\begin{eqnarray}
	L_{\rm CS}[M_S, M_D, y, \cot \theta] \propto
	\theta(\Omega_{\rm obs} - \Omega_{\rm th})
	+\exp\left[-\frac{(\Omega_{\rm th} - \Omega_{\rm obs})^2}{2\,\sigma_{\rm obs}^2}\right]
	\theta(\Omega_{\rm th} - \Omega_{\rm obs})\,,
	\label{eq: LCS}
\end{eqnarray}
where $\Omega_{\rm obs} = 0.1198/h^2$ ($h$ is the normalized Hubble constant) is the cosmological dark matter parameter observed by the PLANCK experiment\,\cite{Ade:2015xua}, while $\sigma_{\rm obs} = 0.0015/h^2$ is the error associated with the observation. We implicitly assume thermal equilibrium abundance of the WIMP as an initial condition, and compute $\Omega_{\rm th}$ using {\tt micrOMEGAs}\,\cite{Belanger:2010gh, Belanger:2008sj, Belanger:2006is} with the input model file for {\tt CalcHEP}\,\cite{Belyaev:2012qa} generated by {\tt FeynRules}\,\cite{Christensen:2008py, Alloul:2013bka}.

The observed dark matter density is considered as an upper bound in the above likelihood function. The most interesting part of the constraint is, of course, the one satisfying the WIMP miracle, namely $\Omega_{\rm th} \simeq \Omega_{\rm obs}$. However, it is often a practice to consider a non-thermal WIMP production in addition to the thermal one. An example is the late time decay of some heavy particle into WIMP, such as a gravitino/moduli decay into a neutralino in the MSSM\,\cite{Moroi:2013sla, Fujii:2001xp}, which gives an additional contribution to the WIMP abundance today. The other example is the late time entropy production, which dilutes the thermally produced WIMP abundance $\Omega_{\rm th}$.\footnote{One should be careful in considering such a scenario as it also dilutes the baryon asymmetry of the universe.} Though such non-thermal mechanisms are not described in the WIMP model, they can exist in some other sectors which do not affect the WIMP phenomenology except the one related to the WIMP abundance.

If we consider the former non-thermal production, the thermally produced abundance $\Omega_{\rm th}$ is required to be less than $\Omega_{\rm obs}$, for the non-thermal production also gives a positive contribution to the WIMP abundance today. The WIMP is then required to have a stronger interaction to the SM particles than the case without the non-thermal production, because $\Omega_{\rm th}$ is inversely proportional to the WIMP annihilation cross-section. On the other hand, if we consider the latter non-thermal production, the opposite situation arises; the WIMP is required to have a weaker interaction. In this paper, we only consider the case with $\Omega_{\rm th} \leq \Omega_{\rm obs}$, allowing the possible existence of the former non-thermal production, because it gives a lower limit on some couplings between the WIMP and the SM particles (while other constraints in the following subsections give upper limits on the couplings) and still allows us to discuss non-trivial WIMP phenomenology including the WIMP miracle region $\Omega_{\rm th} \simeq \Omega_{\rm obs}$.\footnote{If we consider the possible existence of the latter non-thermal production, the WIMP which does not have any interaction to the SM particles at all is allowed without conflicting all the WIMP constraints.} The WIMP is therefore assumed to have the correct relic density $\Omega_{\rm obs}$ in the present universe even if the set of the model parameters gives $\Omega_{\rm th} < \Omega_{\rm obs}$.

In order to obtain the correct dark matter relic density observed today, the WIMP must have interactions with the SM particles with sufficient strength. It usually requires some special mechanisms, and thus gives a stringent constraint on the parameter space. It is not difficult to imagine that the following regions are allowed by the constraint:
\begin{itemize}
	\setlength{\itemsep}{0cm}
	\item Higgs boson and $Z$-boson resonance regions.
	\item Coannihilation region with degenerate $N_1$ and $C$/$N_2/N_3$.
	\item Region in which $N_1$ has a high doublet fraction (smaller $R_S$).
	\item Blind spot region with a large Yukawa coupling $y$.
\end{itemize}
The WIMP annihilation in the resonant regions is enhanced when its mass is close to half of the Higgs or the $Z$-boson mass\,\cite{Mahbubani:2005pt, D'Eramo:2007ga, Calibbi:2015nha, Drees:1992am, Hamaguchi:2015rxa}. In the coannihilation region, there is always a process with a large annihilation cross-section: thanks to the weak charge of a coannihilating particle. This fact is also true for the region in which the WIMP has a large doublet fraction. According to the result of the Higgsino dark matter in the MSSM\,\cite{Mizuta:1992qp, Cirelli:2007xd}, the WIMP mass can be as large as 1\,TeV in both regions. The blind spot region allows us to take large Yukawa couplings while avoiding stringent constraints from the direct dark matter detection experiments. On the other hand, the coupling between the WIMP and the $Z$-boson is not severely constrained at present, for the sensitivity of the detections utilizing a spin-dependent scattering is still low. Thus, the WIMP can efficiently annihilate into top quarks by a $Z$-boson exchange. As a result, the constraint from the dark matter relic density will be avoided even when the WIMP mass is larger than 1\,TeV.

\subsection{Dark matter direct detection}
\label{subsec: DD}

The WIMP is scattered off a nucleon in a spin-independent manner by exchanging a Higgs boson and spin-dependently through a $Z$-boson exchange. Spin-independent WIMP scatterings off a proton and a neutron take place at almost the same rate due to small iso-spin violation, while spin-dependent ones do not. Since no conclusive evidence of a dark matter signal has yet been obtained, we adopt the following likelihood function assuming null signal with its central value being fixed to zero: 
\begin{eqnarray}
	L_{\rm DD}[M_S, M_D, y, \cot \theta] \propto
	\exp\left[-\frac{1}{2}
	\left\{
	\frac{\sigma_{\rm SI}^2}{\delta\sigma_{\rm SI}^2 + \tau_{\rm SI}^2} +
	\frac{\sigma_{\rm SDp}^2}{\delta\sigma_{\rm SDp}^2 + \tau_{\rm SDp}^2} +
	\frac{\sigma_{\rm SDn}^2}{\delta\sigma_{\rm SDn}^2 + \tau_{\rm SDn}^2}
	\right\}
	\right].
\end{eqnarray}
Here, $\sigma_i$ is the scattering cross-section predicted by the model, while $\delta \sigma_i$ is an experimental upper limit at $90\%$\,C.L. divided by $1.64$. Theoretical uncertainty from hadron matrix elements required to compute $\sigma_i$ is also introduced, and is estimated to be $\tau_i = 0.2\,\sigma_i$.\footnote{This uncertainty has been obtained by varying the hadron matrix elements inside {\tt micrOMEGAs}.} Other theoretical uncertainties exist, which are from the local velocity distribution and the local mass density of dark matter. The former one is small compared to those from the matrix elements when the WIMP mass is large enough\,\cite{Catena:2011kv}. On the other hand, the latter one may not be small. Fortunately, all experimental limits are derived assuming that the mass density is 0.3\,GeV/cm$^3$, which is lower than the one obtained using the recent Milky Way mass model\,\cite{Nesti:2013uwa}, and hence we do not take this uncertainty into account.

The most stringent constraint on the spin-independent scattering off a nucleon is from the LUX experiment\,\cite{Akerib:2013tjd}, while those on the spin-dependent scatterings off a proton and a neutron are from PICO-60\,\cite{Amole:2015pla} and LUX\,\cite{Akerib:2016lao} experiments, respectively. The constraints will be updated by the XENON1T experiment\,\cite{Aprile:2012zx, Cushman:2013zza,Aprile:2015uzo} in the near future if no dark matter signals are detected. Moreover, the LZ experiment\,\cite{Malling:2011va, Cushman:2013zza} will eventually update the limits on the spin-independent scattering and the spin-dependent scattering off a neutron, and the PICO250 experiment\,\cite{Cushman:2013zza} will do the same job on the spin-dependent scattering off a proton. We will use these projected limits to investigate how efficiently the future experiments can explore the parameter space of the WIMP model.

The spin-independent and spin-dependent scatterings of the WIMP are controlled by the couplings $y_{11}$ in Eq.\,(\ref{eq: y11}) and $g_{11}$ in Eq.(\ref{eq: g11}), respectively. The coupling $y_{11}$ has already been limited as $y_{11} \leqslant$ 0.043\,(0.076) when the WIMP mass is 300\,GeV\,(1\,TeV). This limit puts a strong constraint on the Yukawa coupling $y$ in general, and, as a result, it forces the coupling $g_{11}$ to be as small as ${\cal O}(10^{-4})$.\footnote{This fact has been numerically confirmed using Monte Carlo data, which we discuss in section\,\ref{sec: results}.} Thus the only possible interactions of the WIMP to the SM particles with sufficient strengths are the weak interactions with the other new particles $C$, $N_2$ and $N_3$, if the WIMP has a high doublet fraction. On the other hand, the severe limit on the coupling $g_{11}$ disappears when the model parameters reside inside the Higgs blind spot region. In this case, even though the present spin-dependent direct detection experiments have directly put a limit on $g_{11}$ as $g_{11} \leqslant$ 0.034\,(0.06) for the WIMP mass of 300\,GeV\,(1\,TeV), the WIMP can still annihilate into top quarks efficiently through a $Z$-boson exchange. This is because its annihilation cross-section is boosted by the longitudinal component of this $Z$-boson such that $\sigma v \simeq 3 g_{11}^2 g^2 m_t^2 / (8\pi c_W^2 m_Z^4)$ with $m_t$ and $m_Z$ being the top quark and the $Z$-boson masses, respectively. Thus, even a WIMP heavier than 1\,TeV can survive in the Higgs blind spot region, as will be shown in the next section. Constraints on the couplings $y_{11}$ and $g_{11}$ will become much more severe: $y_{11} \leqslant$ 0.01\,(0.016), $g_{11} \leqslant$ 0.005\,(0.009) in the near future and $y_{11} \leqslant$ 0.002\,(0.004), $g_{11} \leqslant$ 0.0016\,(0.003) in the future for a WIMP mass of 300\,GeV\,(1\,TeV), if no dark matter signals are detected. Then, the $Z$-boson blind spot region will also be favored due to the direct severe limits on the coupling $g_{11}$, and only the region satisfying both the Higgs and the $Z$-boson blind spot conditions, namely the coannihilation region as shown in the previous section, will survive.

\subsection{Dark matter indirect detection}
\label{subsec: ID}

A number of attempts are now being made to detect dark matter indirectly. Observing cosmic-ray species, such as positrons and anti-protons\,\cite{Adriani:2008zr, Adriani:2010rc, Aguilar:2013qda, Accardo:2014lma, Aguilar:2014mma}, are the well-known ones, where the dark matter signals are expected to be detected as anomalous excesses of cosmic-ray fluxes. The observations are, however, also known to receive large systematic uncertainties originating in the cosmic-ray propagation inside our galaxy and the estimation of the background cosmic-ray fluxes\,\cite{Giesen:2015ufa, Kappl:2015bqa, Evoli:2015vaa}. The uncertainties can be avoided if we utilize neutrinos, even though its detection efficiency is still too low to put a strong limit\,\cite{Abbasi:2011eq, Aartsen:2013dxa, Aartsen:2014hva, Aartsen:2015xej}. On the other hand, observing gamma-rays, in particular those from Milky Way satellites called dwarf spheroidal galaxies (dSphs), are regarded as the most effective way to detect dark matter\,\cite{Ahnen:2016qkx, Ackermann:2015zua, Drlica-Wagner:2015xua}. The systematic uncertainties associated with a dark matter distribution inside each dSph, however, still remains\,\cite{Martinez:2013els, GeringerSameth:2011iw, Bonnivard:2015vua, Bonnivard:2015xpq, Bonnivard:2015tta, Ullio:2016kvy, Hayashi:2016kcy}, so that its detection capability at present is not strong enough if we take these into account.

Among several indirect dark matter detections, the one utilizing the cosmic microwave background (CMB) currently allows us to put a robust limit on dark matter annihilation\,\cite{Chen:2003gz, Padmanabhan:2005es, Zhang:2006fr, Mapelli:2006ej, Slatyer:2009yq, Galli:2009zc, Cirelli:2009bb, Slatyer:2015jla, Kawasaki:2015peu, Slatyer:2015kla, Kanzaki:2009hf, Galli:2013dna}. This is because the universe at the recombination epoch is described by the linear density perturbation, so that all systematic uncertainties mentioned above, which are caused mainly by the non-linearity of the perturbation, can be avoided. The universe at the epoch was made up of the thermal plasma among photons, electrons, protons, neutral hydrogen and helium. An efficient annihilation of dark matter injecting energetic particles into the plasma affects the recombination history of the universe, resulting in the boost of the residual ionization fraction.\footnote{The dark matter annihilation also affects the thermal history of the Big Bang nucleosynthesis\,\cite{Kawasaki:2015yya}.} This effect is imprinted in the CMB spectrum, and precise observations enable us to detect the effect even if it is caused by a small change of the fraction. Since such an effect has not yet been detected, we put a constraint on the singlet-doublets WIMP model by adopting the following likelihood function:
\begin{eqnarray}
	L_{\rm ID}[M_S, M_D, y, \cot \theta] \propto
	\theta[{\rm UL}\,(M_{N1}) - \langle \sigma v \rangle],
\end{eqnarray}
where $\theta[x]$ is the Heaviside step function, while $\langle \sigma v \rangle$ is the annihilation cross-section predicted by the model, which is evaluated by taking its thermal average at the temperature of the recombination epoch. The upper limit on the cross-section at $90\%$ confidence level is obtained as ${\rm UL}\,(M_{N1}) = 1.453 \times 10^{-27}\,(M_{N1}/1\,{\rm GeV})^{1.05}$\,cm$^3$/s\,\cite{Slatyer:2015jla} based on the PLANCK result.\,\cite{Ade:2015xua}. This limit will give a constraint on the parameter region with a light WIMP having a high doublet fraction, because such a WIMP has a large self-annihilation cross-section with a large enough number density at the recombination epoch.

The constraint on the annihilation cross-section will be upgraded by several factors in the future\,\cite{Madhavacheril:2013cna}, and a heavier WIMP with a high doublet fraction can thus be explored. WIMP with a mass of ${\cal O}(1)$\,TeV seems, however, difficult to be searched for by this detection. Such a heavy WIMP region can potentially be covered by the indirect detection utilizing gamma-ray observations of dSphs when dark matter profiles inside dSphs are precisely evaluated by accumulating enough kinematical data of the galaxies\,\cite{Bhattacherjee:2014dya}.

\subsection{Dark matter searches at colliders}
\label{subsec: CL}

Recent and future colliders, such as the Large Electron Positron Collider (LEP), the Tevatron, the LHC, the International Linear Collider (ILC)\,\cite{Behnke:2013xla}, the Circular Electron Positron Collider (CEPC)\,\cite{CEPC-SPPCStudyGroup:2015csa} and the Future Circular Collider of Electrons and Positrons (FCC-ee)\,\cite{Gomez-Ceballos:2013zzn}, are designed to look for signatures of physics beyond the standard model (BSM). The LEP and Tevatron have not detected any conclusive BSM signatures. The LHC is now on a hunt for such signatures, and the hunt will be the most important task at future colliders (ILC, CEPC and FCC-ee). We consider several constraints on the singlet-doublets WIMP model obtained by collider experiments performed so far, and discuss some prospects of looking at the signatures of this model in the (near) future.

\subsubsection{Invisible Z-boson decay}

When $m_{N1} < m_Z/2$, the $Z$-boson can decay into a pair of WIMPs. This process should be regarded as a part of the invisible decay width of the $Z$-boson ($\Gamma_Z^{\rm Inv}$) at collider experiments. The measurement of the total decay width of the $Z$-boson ($\Gamma_Z^{\rm Tot}$) at the LEP was very precise, so that it gives an upper limit on the width as $\Gamma_{Z}^{\rm Inv} \leqslant$ 2\,MeV at 90\% confidence level\,\cite{ALEPH:2005ab}. We thus adopt the following likelihood function to involve this limit:
\begin{eqnarray}
	L_{\rm InvZ}[M_S, M_D, y, \cot \theta] =
	\exp\left[-\frac{(\Gamma_Z^{\rm Inv})^2}{2\,(2\,{\rm MeV}/1.64)^2}\right],
\end{eqnarray}
where the invisible decay width $\Gamma_Z^{\rm Inv}$ is computed within the singlet-doublets WIMP model. At first glance, this constraint seems very severe and puts a stringent limit on the model parameter space, in particular, on the parameter region where the WIMP coupling to the $Z$-boson is not suppressed. This is, however, not true, for spin-dependent direct dark matter detections are already putting more stringent limit on the coupling, and this trend will be more strengthened in the (near) future if no WIMP signal is detected. On the other hand, when the WIMP is much lighter than 10\,GeV, the sensitivities of the direct detection experiments are very weak, so that the invisible width search can play a dominant role to put limits on the parameter space. Such a light WIMP region is, however, being limited by other dark matter searches: the new particle searches at the LEP already kill the region when the WIMP has a significant doublet fraction, while the WIMP annihilation cross-section in the region is too small to satisfy the constraint from the dark matter relic abundance when the WIMP is close to an 
SU(2)$_L$ singlet.

\subsubsection{Invisible Higgs decay}

When the WIMP is lighter than half of the Higgs mass, the Higgs boson can decay into a pair of the WIMPs, which is, this time, regarded as the invisible decay width of the Higgs boson ($\Gamma_h^{\rm Inv}$). The invisible decay branching ratio is being constrained by a global fit of Higgs data at the LHC, which leads to an upper limit on the ratio as ${\rm Br}\,(h \to N_1 N_1) \leq 0.24$ at 90\% confidence level\,\cite{Giardino:2013bma}. We thus adopt the following likelihood function:
\begin{eqnarray}
	L_{\rm InvH}[M_S, M_D, y, \cot \theta] =
	\exp\left[-\frac{{\rm Br}\,(h \to N_1 N_1)^2}{2\,(0.24/1.64)^2}\right].
\end{eqnarray}
Using the invisible decay width $\Gamma_h^{\rm Inv}$ computed from the singlet-doublets WIMP model, the branching ratio is defined as ${\rm Br}\,(h \to N_1 N_1) \equiv \Gamma_h^{\rm Inv}/[\Gamma^{\rm SM}_h + \Gamma_h^{\rm Inv}]$, where the total decay width of the Higgs boson within the SM framework is denoted by $\Gamma^{\rm SM}_h \simeq 4.08$\,MeV\,\cite{Heinemeyer:2013tqa} with the Higgs boson mass being $m_h = 125.09$\,GeV\,\cite{Aad:2015zhl}. This constraint puts a limit on the WIMP coupling to the Higgs boson. However, as in the case of the $Z$-boson invisible width, this coupling has already been severely constrained by spin-independent direct dark matter detections. As a result, the invisible Higgs width measurement will not play an important role to put a limit on the model parameter space.

\subsubsection{Electroweak precision measurements}

We inevitably introduce particles having electroweak interactions in the singlet-doublets WIMP model, so that oblique corrections\,\cite{Peskin:1991sw} might be a useful tool to detect such a WIMP indirectly even if it cannot be produced directly at lepton colliders (see the next subsubsection for more details). As is usual, the so-called $S$ and $T$ parameters can play important roles in the WIMP model: the $S$ parameter measures the size of the weak isospin breaking but it turns out to give a loose constraint on the model due to small field degrees of freedom. On the other hand, a severe constraint may come from the $T$ parameter, which measures the size of the custodial symmetry breaking, because it is proportional to the difference of the Yukawa couplings $y_1^2 - y_2^2$ and can thus be sizable for a larger $y$ and a smaller $ \cot \theta$. With the use of Monte Carlo data discussed in the next section, the $T$ parameter, however, turns out to be always smaller than 0.1 after all the other constraints are imposed. This value is well below the current upper bound on the $T$ parameter\,\cite{Baak:2014ora}.

As was mentioned in section\,\ref{subsec: DD}, if no new physics signals are detected, all the parameter regions of the WIMP model will shrink into the coannihilation region in the (near) future due to severe constraints from direct dark matter detections. In fact, it leads to a larger $\cot \theta$ for a larger Yukawa coupling $y$. As a result, contributions to the $T$ parameter from the WIMP is less and less significant. After all, observing the oblique corrections, namely the electroweak precision measurements, is not (will not be) an ideal way to detect the WIMP at present (future), and hence we do not include it in our analysis.

\subsubsection{New particle searches at lepton colliders}

New particles having electroweak interactions can be easily searched for at lepton colliders. The singlet-doublets WIMP model predicts an electrically charged (Dirac) particle $C^\pm$ and three electrically neutral (Majorana) particles $N_1$, $N_2$ and $N_3$.

The charged particle $C$ can be pair produced at the colliders by exchanging a photon and a $Z$-boson in the $s$-channel, and its annihilation cross-section is always large thanks to its electroweak quantum numbers. The charged particle can therefore be detected irrespective of its decay mode whenever its production is kinematically allowed.\footnote{Mono-photon signals can be expected even if the mass difference $m_C - m_{N_1}$ is small\,\cite{Abdallah:2003xe, Chen:1995yu}.} Since conclusive signals of new particle productions were not obtained at the LEP, we have a lower limit on the mass of the charged particle $M_D$. Thus in the next section, we perform a numerical scan of the model parameter space in the range {\boldmath \bf $M_D \geq$ 100\,GeV}\,\cite{Abdallah:2003xe}.

On the other hand, the neutral particles are pair produced by exchanging only a $Z$-boson in the $s$-channel at lepton ($e^+e^-$) colliders. Its signal strength thus depends on how a large coupling to the $Z$-boson the neutral particle pair has. Since the charged particle production has already put a constraint on the parameter space, it is expected that the neutral particle productions in the singlet-doublets WIMP model can be important in the following three cases: First one is that all charged and neutral particles are light enough to be produced at the colliders. The charged particle pair production however puts a more stringent constraint than those of the neutral particle pair. Next one is that doublet-like neutral particles are lighter than the singlet one. It however leads to the same result as the previous case, for the charged particle is also as light as the light neutral particles. The last case is opposite to the second case, where a singlet neutral particle is lighter than the doublet ones. Since the $N_1$-$N_1$-$Z$ coupling is already constrained by the spin-dependent direct dark matter detection experiments, the relevant process is $e^-e^+ \to N_1 N_2 (N_3)$. The coupling of $N_1$ to the $Z$-boson with $N_2$ or $N_3$ is, however, suppressed in this case because of the singlet nature of $N_1$
, so that the LEP experiment does not put any significant limit on the model. There may be a more subtle spectrum for the neutral particles. It however turns out that by generating Monte Carlo data as shown in the next section, all the cross-sections of the neutral particle pair productions are always below 1\,pb after imposing all other constraints. Thus, the LEP experiment does not put any constraint on the neutral particle pair production in the singlet-doublet WIMP model, and hence we do not include it in our analysis.

Though lepton colliders do not play an important role to put a constraint on the singlet-doublets WIMP model at present, these can potentially be important in the (near) future. As was already mentioned in previous sections, if no WIMP signals are detected in the (near) future from all kinds of dark matter search experiments, then only the coannihilation region remains as the unexplored one with the WIMP mass being larger than a few hundred GeV. The WIMP in such a region can hardly be detected at both direct and indirect dark matter detections, while future lepton colliders such as the ILC, CEPC and FCC-ee can search for it at least through the charged particle pair production if the center of mass energy of the collisions are large enough.

\subsubsection{New particle searches at hadron colliders}
\label{subsubsec: hadron colliders}

New particles having electroweak interactions are also searched for at the LHC. One of the prominent examples is the pair production of charginos and neutralinos in the framework of supersymmetry. This search can be recast to the singlet-doublets WIMP model thanks to the analogy between this WIMP model and the MSSM, and thus can put constraint on the parameter space. Among various pair production channels, that of a charged particle $C$ and a heavier neutral particle $N_2$ or $N_3$ seems the best one, for it gives a rather clean signal, namely the three leptons plus missing transverse energy final state, even in the messy environment of the hadron collider. The full process reads
\begin{equation}
	p p \to C\,N_{2,\,3} \to (W^{(*)}\,N_1)\,(Z^{(*)}\,N_1) \to 3 \ell + \slashed{E}_T.
\end{equation}
Here, the superscript `$(*)$' indicates that the $W$ and $Z$-bosons can go off-shell as well. We include the off-shell $W$ and $Z$-bosons as well because we hope to have lesser backgrounds in this region since the on-shell $Z$-production can be vetoed away.

\begin{figure}[t]
\centering
\includegraphics[width=0.7\textwidth]{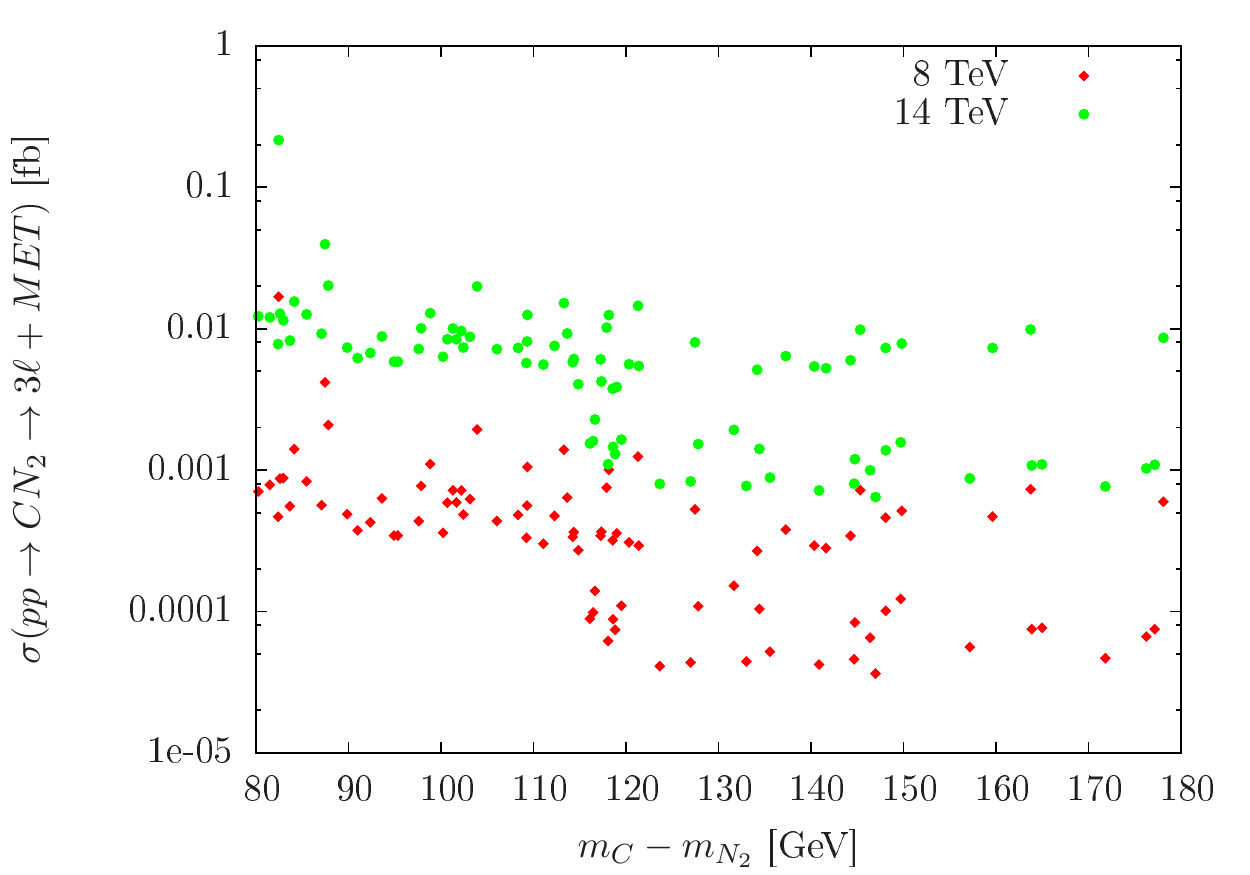}
\caption{\sl \small Parton level cross-section for $p p \to C N_2 \to 3 \ell + \slashed{E}_T$}
\label{fig:3lMET-eps-converted-to.pdf}
\end{figure}

It has been shown in Ref.\,\cite{Calibbi:2015nha} that the corresponding ATLAS search at 8\,TeV\,\cite{Aad:2014nua} is capable of covering the region, $m_{N_2}$, $m_{N_3}$, $m_C$ $\lesssim$ 270\,GeV and $m_{N_1} \lesssim$ 75\,GeV, assuming that the average of the branching fractions Br$\,(N_2 \to Z N_1)$ and Br\,$(N_3 \to Z N_1)$ is at least 60\%. We thus perform a thorough study to investigate how this search puts constraints on our available model parameter space. However, it is shown not to have any significant power to rule out the parameter space that has already been limited by other dark matter constraints. For the current model, the charged particle, $C$ will always decay to one of the neutral particles and 
an on- or off-shell $W$ boson because these are the dominant decay modes. The branching ratios are 
largely independent of the mass differences between the charged and neutral particles. However, for a small
mass difference, the cut efficiencies for various cuts will begin to decrease because of a dearth of phase space, for
instance the lepton $p_T$ will peak at smaller values and hence the $p_T(\ell_1,\ell_2,\ell_3) > 50$ GeV cut 
will yield a considerably smaller efficiency. After computing the $3 \ell + \slashed{E}_T$ cross-section, we find that 
this is not enough to put any significant constraints on the parameter space as the cross-section varies between 
$10^{-4}$ fb and a few times $10^{-3}$ fb before any detector level cuts. The $C N_2$ pair production cross-section can however be as large as 0.1 fb at 8 TeV (and can be as large as 1 fb for the 14 TeV run) but the branching ratio suppresses the final cross-sections to significantly smaller values. We must also note that, we are working in the parameter region allowed by XENON1T. Hence we find that after implementing {\tt CheckMATE}\,\cite{Drees:2013wra, deFavereau:2013fsa, Cacciari:2011ma, Cacciari:2008gp, Read:2002hq} for this ATLAS analysis, one cannot exclude a single point of the parameter space allowed by other constraints. The ChecKMATE analysis includes all the five main signal regions and the sub-regions which have been optimised following Ref.~\cite{Aad:2014nua}. For completeness, we include a plot (Fig.~\ref{fig:3lMET-eps-converted-to.pdf}) showing the signal cross-section in the $3 \ell + \slashed{E}_T$ final state as a function of $m_C-m_{N_2}$ for the 8 TeV and 14 TeV cases.

As mentioned before, the $3 \ell +\slashed{E}_T$ cross-section in the parameter region allowed by the direct
detection constraints in the (near) future, will be suppressed mainly because the average mass of the particles $C$ and
$N_{2,3}$ is large. Also when they are nearly degenerate with $N_1$, the cut-efficiency of the signal is reduced because
of softer leptons produced from off-shell $W$ bosons and hence unable to satisfy certain cuts demanding larger values for
the $p_T$ of the leptons. We also perform ATLAS's prediction study for the high-luminosity run at 14\,TeV (See Appendix\,\ref{app: lhc14}). However it is also shown not to have any significant power to exclude the parameter space.\footnote{
Ref.\,\cite{Hamaguchi:2015rxa} plays a complementary role to our study, which also includes the singlet-like regime with $R_s < 0.05$. It has been shown that, for the case of the singlet-like regime with a supersymmetric $Z$- and Higgs-resonant neutralino dark matter, the ATLAS search at 14\,TeV\,\cite{ATL-PHYS-PUB-2014-010} can be a strong probe. In another complementary study, Ref.~\cite{vanBeekveld:2016hbo} has shown in the context of pMSSM that the $3\ell + \slashed{E}_T$ channel can have constraining power for $m_{\textrm{NLSP}} - m_{\textrm{LSP}} \sim (5-50)$ GeV.
} With the proposed 33\,TeV High-energy LHC (HE-LHC) or the 100\,TeV Very Large Hadron Collider (VLHC) in the future, the signal cross-sections can be enhanced. However, one has to deal with very large backgrounds and a full analysis at such colliders is beyond the scope of this paper. Hence, at least for the singlet-doublets WIMP model, the direct detection constraints play a stronger role than those from the LHC, and we do not consider constraints from new particle searches at the LHC in our analysis.

\section{Scanning results}
\label{sec: results}

We are now in a position to present our scanning results to quantitatively display which parameter regions of the singlet-doublets WIMP model survive at present, will be covered in the near future, and will be left over in the future. The strategy of our numerical scanning is as follows: Engaging with {\tt MultiNest\,v2.18}\,\cite{Feroz:2008xx} of 20000 living points, a stop tolerance factor of $10^{-4}$, and an enlargement factor (reduction) parameter of 0.8, we perform several random scans in the following parameter space, 
\begin{eqnarray}
	10\,{\rm GeV} & \leq\,\,\,\,M_S\,\,\,\,\leq & 2\,{\rm TeV}, \nonumber \\
	100\,{\rm GeV} & \leq\,\,|M_D|\,\,\leq & 2\,{\rm TeV}, \nonumber \\
	0 & \leq\,\,\,\,\,\,y\,\,\,\,\,\,\leq &1, \nonumber\\
	0 & \leq\,\cot\theta\,\leq & 1.
	\label{eq: domain}
\end{eqnarray}
The doublet mass parameter $M_D$ can either be positive or negative. Some of the regions such as the blind spot regions or the resonance regions have very low prior probabilities to be accessed if only flat or log prior distributions are adopted. To obtain a better coverage, similar to a Bayesian approach, we update our priors, namely, we start with one log and one flat prior scans, but later adjust our scan range and prior distribution based on the previous likelihood distribution. In the end, we obtain in total more than $10^6$ points out of which for only $6 \times 10^5$ points $\chi^2 \equiv -2\ln\,L[M_S, M_D, y, \cot\theta] < 50$ have been recorded and used for our figures. Our minimal $\chi^2$ value turns out to be almost zero. Below, we present our scanning results for the present, the near future, and the future.

\subsection{Present status}
\label{subsec: present}

\begin{figure}[t]
	\centering
	\subfloat[\label{subfig: MN_MD_P}] { \includegraphics[width=0.48\textwidth]{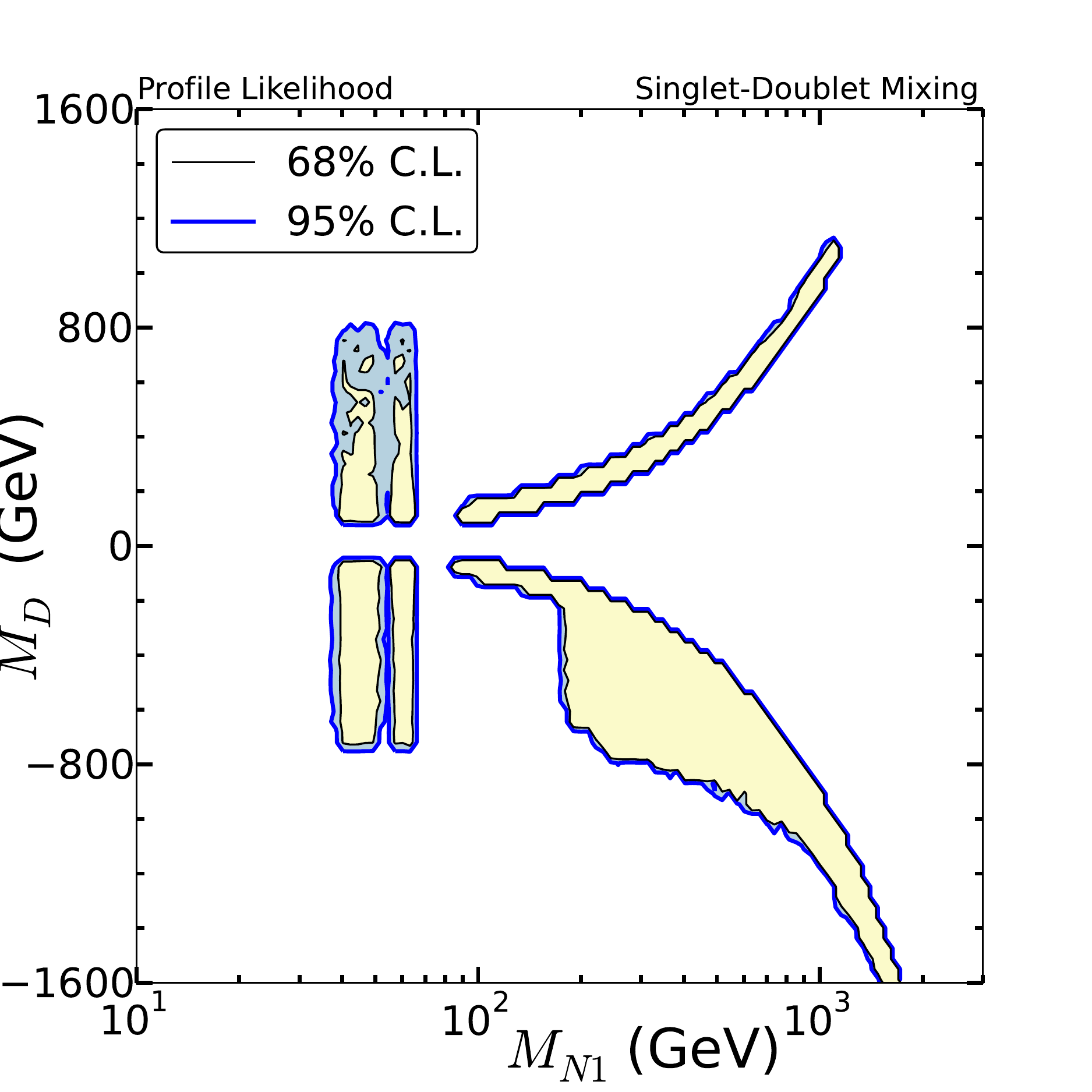} }
	\subfloat[\label{subfig: MN_RS_P}] { \includegraphics[width=0.48\textwidth]{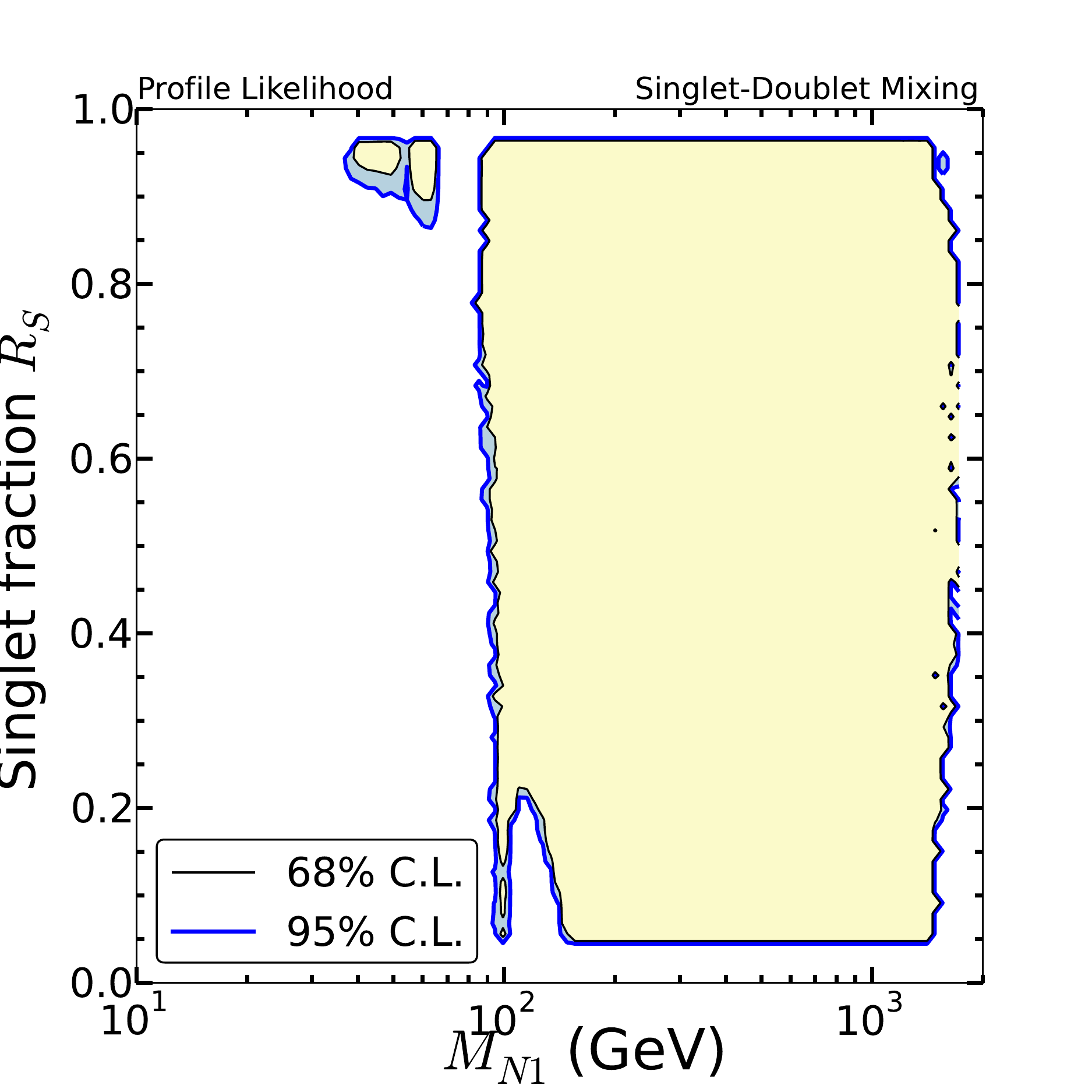} }
	\caption{\sl \small Present constraints on the WIMP parameter space in the singlet-doublets WIMP model.}
	\label{fig: present}
\end{figure}

The present status of the singlet-doublets WIMP model can be seen in Fig.\,\ref{fig: present}, where the 1$\sigma$\,(2$\sigma$) contour is shown with the yellow (blue) band. The left and right panels (Figs.\,\ref{subfig: MN_MD_P} and \ref{subfig: MN_RS_P}) show a viable model parameter space at present in the planes of $(M_{N1}, M_D)$ and $(M_{N1}, R_S)$, respectively (Comprehensive figures showing the whole viable parameter space after imposing present experimental constraints are provided in Appendix\,\ref{app: figures}.). The most important feature is an asymmetry in Fig.\,\ref{subfig: MN_MD_P} with respect to the sign of the doublet mass parameter, $M_D \leftrightarrow - M_D$, owing to the stringent constraints form spin-independent direct dark matter detections. As was mentioned in the previous section, these constraints can be avoided if the model parameters lie in the Higgs blind spot region, and the regions appears only when $M_D < 0$. Below, we will look into the figure in more details.

The mass parameter $M_D$, whose absolute value gives the mass of the new charged particle $C$, is limited to be $|M_D| > 100$\,GeV due to the LEP constraint. There are two regions when $M_D > 100$\,GeV. One of these comes from the coannihilation region with $M_{N1} \sim M_D$. This mass degeneracy between the WIMP and the particle $C$ is required to satisfy the constraint from the dark matter relic abundance and also the well-tempered condition in Eq.\,\eqref{eq: well-tempered condition}, because the size of the Yukawa couplings are always limited when $M_D > 100$\,GeV. The upper bound on the mass parameter $M_D \lesssim 1$\,TeV in this coannihilation region can be understood by an analogy with the Higgsino-like dark matter in the MSSM. The other one is the Higgs/$Z$-boson resonance region with $M_{N1} < 100$\,GeV. The WIMP here is singlet-like because $M_D > 100$\,GeV, as can also be seen in Fig.\,\ref{subfig: MN_RS_P}. The upper bound on the mass parameter $M_D \lesssim 800$\,GeV comes from the well-tempered condition given in Eq.\,\eqref{eq: well-tempered condition}.

On the other hand, an additional parameter region becomes available when $M_D < -100$\,GeV, owing to the Higgs blind spot condition, which allows large Yukawa couplings. This region appears when $M_{N1}$ is larger than the top quark mass, where the annihilation of the WIMP into a pair of top quarks is boosted by the longitudinal component of the $s$-channel exchanged $Z$-boson, as discussed in Sec.\,\ref{subsec: DD}. It also takes a role to relax the mass degeneracy between the WIMP and the charged particle $C$. Here also, the lower bound on $M_D$ comes from the condition in Eq.\,\eqref{eq: well-tempered condition}. In the resonance region with $M_D < -100$\,GeV, a slightly lighter WIMP mass is allowed compared to the one with $M_D > 100$\,GeV, because larger Yukawa couplings help the total WIMP annihilation to be sufficiently large, opening up a lighter WIMP mass region slightly away from the pole.

Here, we also comment on the role of indirect dark matter detections. As can be seen in Fig.\,\ref{subfig: MN_RS_P}, the constraint from the CMB observation excludes some parameter regions; a characteristic spike around $M_{N1} \sim 120$\,GeV and $R_S \lesssim 0.2$. This fact means that the indirect detection does not play an important role for the WIMP when $R_S$ larger than 0.2. This is because the s-channel (velocity unsuppressed) annihilation of the well-tempered WIMP into SM particles (e.g. weak gauge bosons) is suppressed when $R_S$ increases. Moreover, even if $R_S$ is small enough, the constraint becomes less significant when the WIMP is heavier, for the number density of the WIMP in the present universe decreases.\footnote{The spike structure of the region with $M_{N1} \sim 120$\,GeV and $R_S \lesssim 0.2$ is because of this reason and the threshold behavior of the s-channel (velocity unsuppressed) annihilation process, $N_1 N_1 \to W^+ W^-$.}

\subsection{Near future prospects}
\label{subsec: near future}

\begin{figure}[t]
	\centering
	\subfloat[\label{subfig: MN_MD_NF}] { \includegraphics[width=0.48\textwidth]{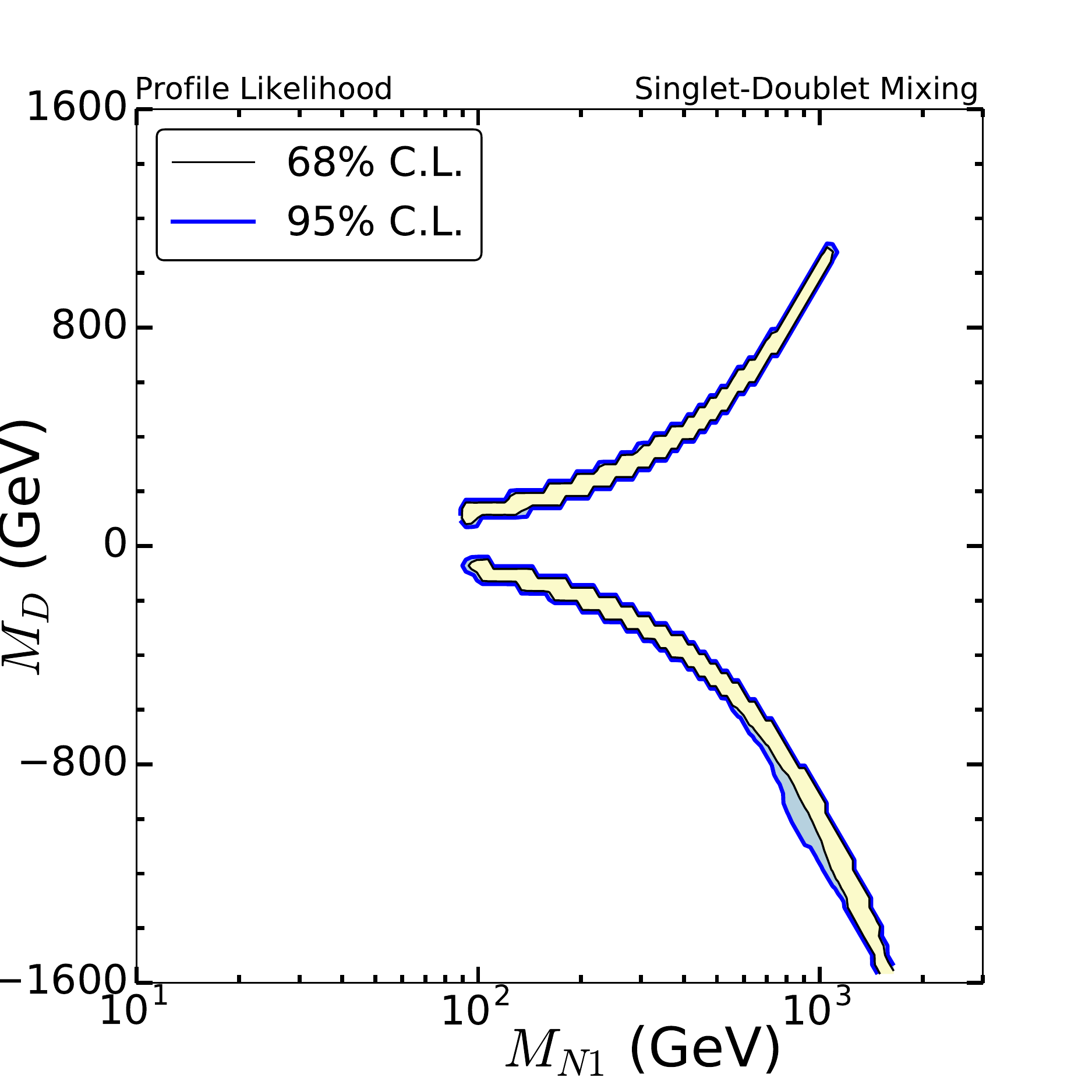} }
	\subfloat[\label{subfig: MN_RS_NF}] { \includegraphics[width=0.48\textwidth]{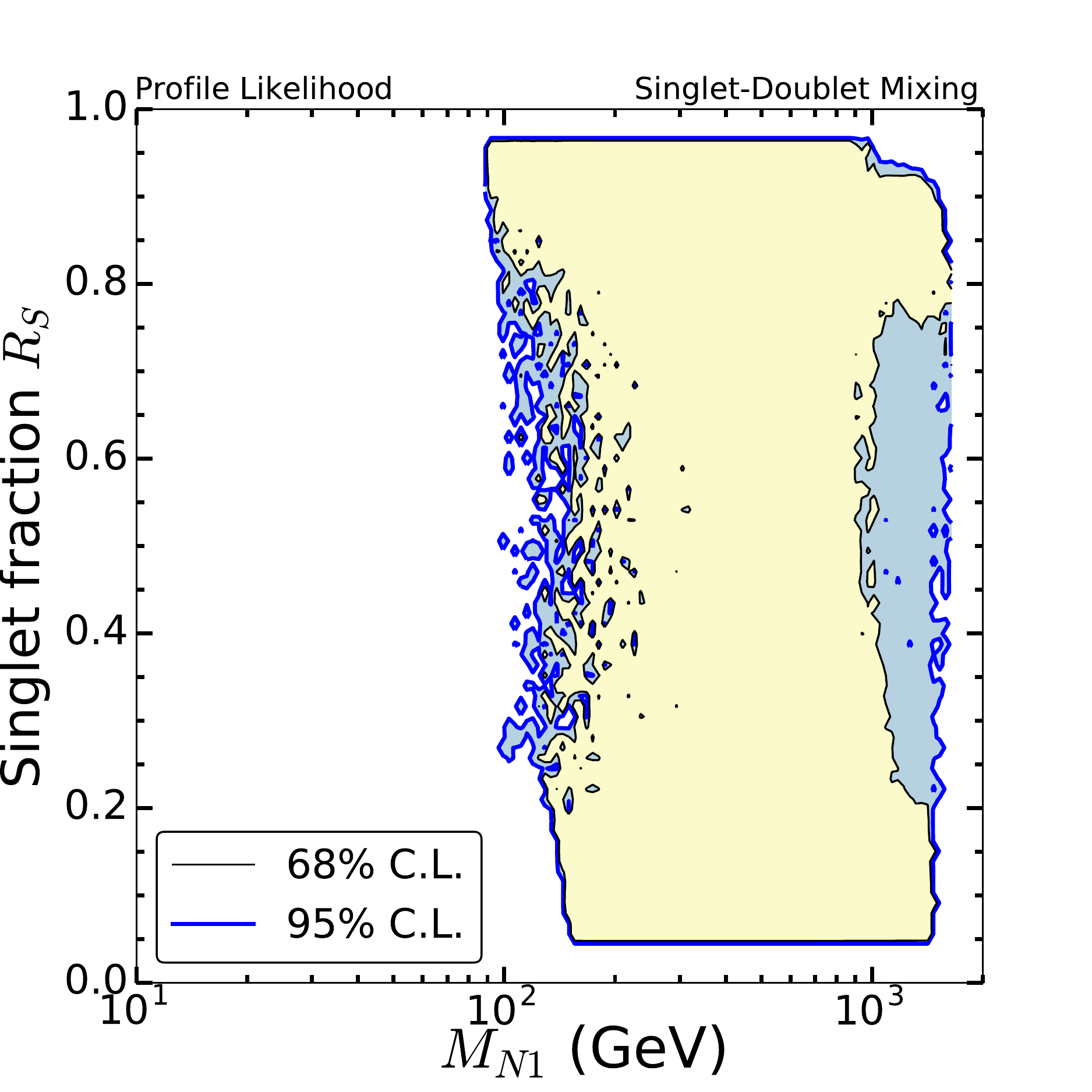} }
	\caption{\sl \small Potential constraints on the WIMP parameter space in the near future in the singlet-doublets WIMP model.}
	\label{fig: near future}
\end{figure}

The expected viable model parameter space after the XENON1T experiment is shown in Fig.\,\ref{fig: near future}, assuming that the experiment will not observe any dark matter signals (Comprehensive figures showing whole viable parameter space after imposing the XENON1T constraints are provided in Appendix\,\ref{app: figures}.). The crucial difference from the present status in Fig.\,\ref{fig: present} is that the Higgs/$Z$-boson resonance region and the Higgs blind spot region seem to disappear in Fig\,\ref{subfig: MN_MD_NF}. This is because not only the spin-independent but also the spin-dependent direct dark matter detections will be much improved as mentioned in Sec.\,\ref{subsec: DD}, and hence these will put very severe limits on the singlet-doublets WIMP model. As a result, only the coannihilation regions in both $M_D > 100$\,GeV and $M_D < -100$\,GeV seem to survive. Note that even if the blind spot region disappears, the coannihilation region, which is realized in both $M_D > 100$\,GeV and $M_D < -100$\,GeV regions, can survive, for the coannihilating particle such as $C$ has a large enough annihilation cross-section, as discussed in Sec.\,\ref{subsec: CS}.

The importance of the spin-dependent direct dark matter detection can be seen from Fig.\,\ref{subfig: MN_RS_NF}. It is indeed confirmed that the resonant region will be excluded. On the other hand, the region with $M_{N1} \gtrsim 1$\,TeV still survives. The region with such a large WIMP mass is realized when the coupling between the WIMP and the $Z$-boson ($g_{11}$) is large enough, and such a large coupling is realized only in the Higgs blind spot region. This fact means that some part of the blind spot regions turns out to survive yet in the near future even if any WIMP signal is not detected until then. Because the coupling $g_{11}$ can be directly constrained by the spin-dependent cross-section measurement, the region will be eventually excluded in future if no WIMP signal is detected. (See also Fig.\,\ref{appfig: Y_COT_NF}, where the region with large Yukawa couplings $y$ is pushed to the $Z$-boson blind spot region, namely $\cot \theta \to1$.).

Here, it is also worth pointing out that all the new particles are degenerate in the coannihilation region, and hence it is very hard to discover their signals at hadron colliders due to the small mass splitting between $N_1$ and $C/N_{2,3}$, as discussed in Sec.\,\ref{subsubsec: hadron colliders}.

\subsection{Future prospects}
\label{subsec: future prospects}

\begin{figure}[t]
	\centering
	\subfloat[\label{subfig: MN_MD_F}] { \includegraphics[width=0.48\textwidth]{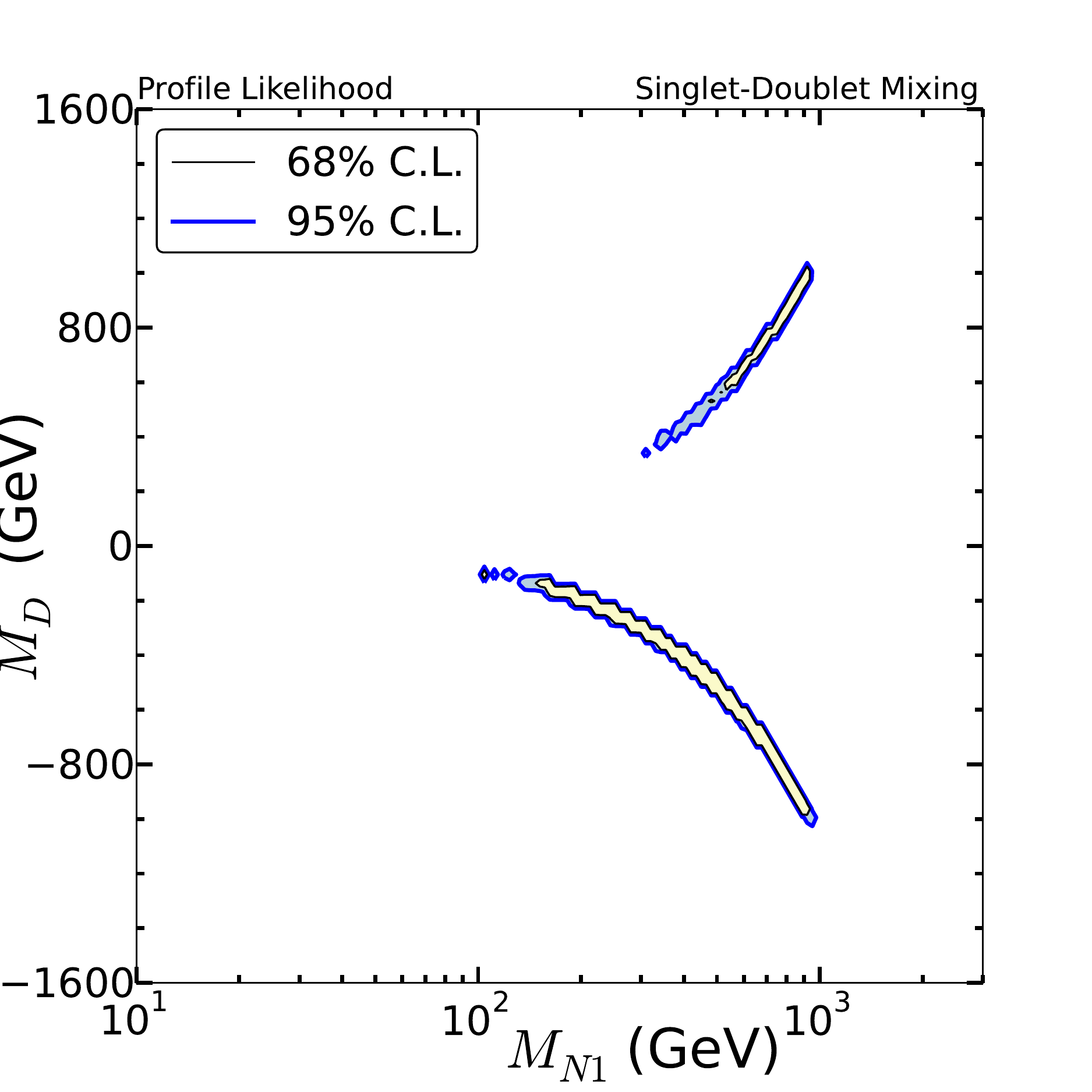} }
	\subfloat[\label{subfig: MN_RS_F}] { \includegraphics[width=0.48\textwidth]{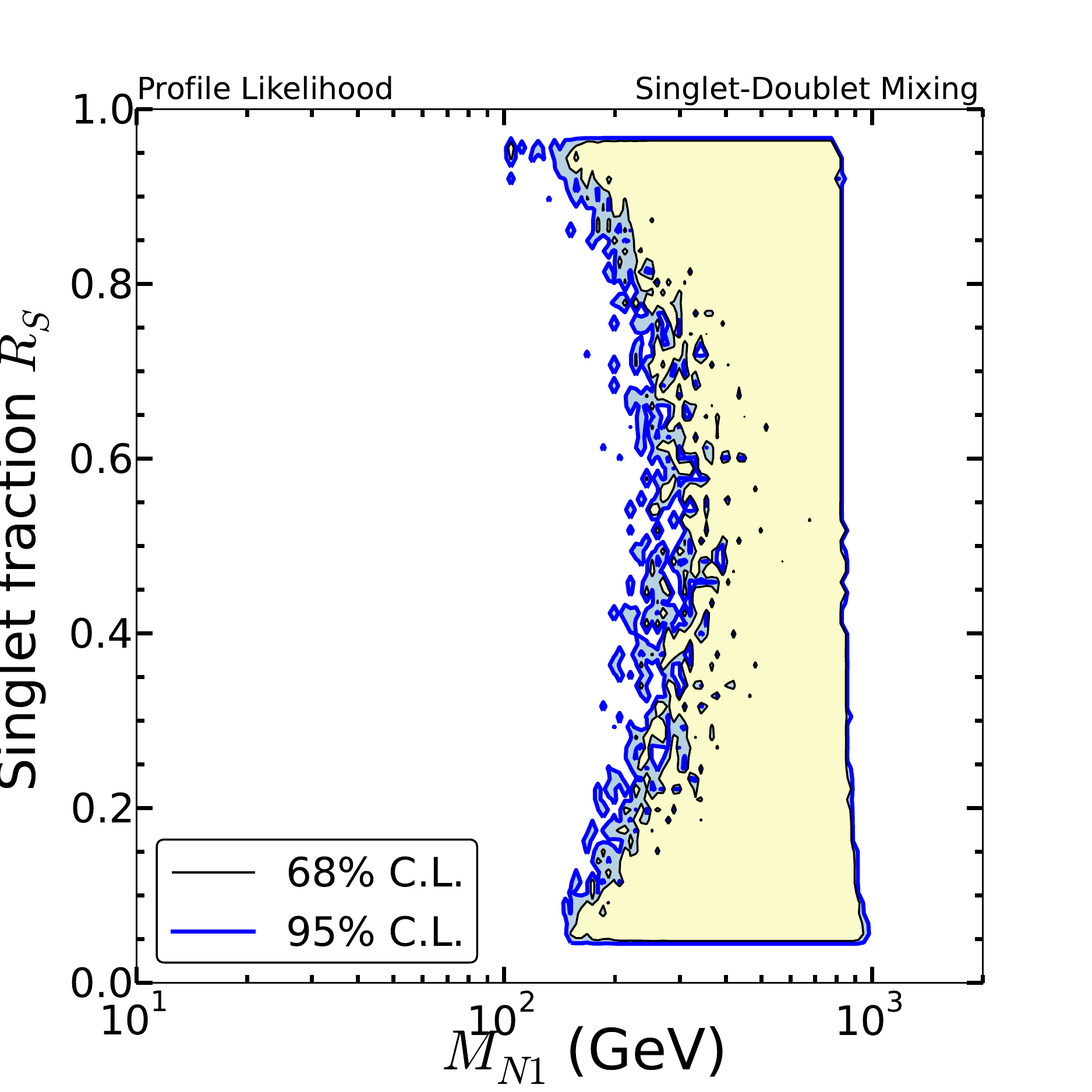} }
	\caption{\sl \small Future prospects on the WIMP in the singlet-doublets WIMP model.}
	\label{fig: future}
\end{figure}

Future prospects on the singlet-doublets WIMP model after the LZ and the PICO250 experiments are shown in Fig.\,\ref{fig: future}. The properties look similar to those from the near future constraints, even though the constraints from the spin-independent and the spin-dependent direct dark matter detections are much more severe. As a result, almost all the viable model parameter space shrinks to the coannihilation regions. The remaining region with $M_D < 100$\,GeV is the one which almost satisfies both the Higgs and the $Z$-boson blind spot conditions simultaneously; $M_{N1} \simeq M_S \simeq - M_D$ and $\cot \theta \simeq 1$ (See also Fig.\,\ref{appfig: future} in Appendix\,\ref{app: figures} for further confirmation.). It can also be seen from Fig.\,\ref{subfig: MN_MD_F} that the small coannihilation region with $M_D > 100$\,GeV, which is outside the Higgs-blind spot, will survive, even though an extreme tuning of the mass splitting is required. Therefore, in the future, the singlet-doublets WIMP model will be constrained alone by the future direct dark matter detections to the corner of the coannihilation region, if no signals are detected.

\section{Summary and Discussion}
\label{sec: summary}

We have investigated the current status and the (near) future prospects of a fermionic WIMP in the well-tempered regime, particularly focusing on the simplest case of the minimal composition setup, namely the singlet-doublets WIMP model. It then turns out that a viable model parameter space at present can be classified into the following regions:
\begin{itemize}
	\setlength{\itemsep}{0cm}
	\item[(i)] Higgs boson and $Z$-boson resonance regions.
	\item[(ii)] Coannihilation region with degenerate $N_1$ and $C$/$N_2/N_3$.
	\item[(iii)] Region in which $N_1$ has a high doublet fraction (smaller $R_S$).
	\item[(iv)] Blind spot region with a large Yukawa coupling $y$.
\end{itemize}
Owing to the stringent constraint from the LUX experiment on the spin-independent cross-section for a WIMP scattering, the coupling between the WIMP and the Higgs boson is already constrained, and hence it is pushing the viable model parameter space towards the Higgs-blind spot region, even though other regions still survive. A low mass region with a high doublet fraction is somewhat constrained by the CMB measurement, which was adopted as a robust indirect dark matter constraint in our analysis.

In the near future, significant improvements on the spin-independent and the spin-dependent WIMP scattering cross-sections are expected from the XENON1T experiment. The WIMP couplings to both the Higgs and the $Z$-bosons will be severely constrained if the experiment does not observe any dark matter signals. Then, the viable model parameter region tends to shrink to those regions where both the Higgs and the $Z$-boson blind spot criteria are satisfied. This trend will be strengthened in the future, when the LZ and the PICO250 experiments update their constraints. The viable parameter space will then shrink to the one where the regions (ii) and (iv) overlap, or the small one where only region (ii) is satisfied without satisfying the Higgs blind spot condition.

It is important to consider what kinds of experiments have sufficient capability to cover the leftover regions, assuming that none of the future direct dark matter detection experiments unfortunately observe any dark matter signals. Hadron colliders such as the LHC seem difficult to do it even if their luminosities are high enough, for all the new particles are highly degenerate in mass in the leftover regions. The WIMP in these regions may have a significant doublet fraction as can be seen from Fig.\,\ref{subfig: MN_RS_F}. Indirect dark matter detections may probe such WIMPs if systematic errors associated with astrophysical uncertainties are within good control. However, if the WIMP has a large singlet fraction, the indirect detections do not work at all, for the WIMP annihilation is severely $p$-wave suppressed in the present universe. In such a case, future lepton colliders such as the ILC can be useful in exploring the leftover region directly (via the pair production of the charged particle $C$, etc.) and indirectly (via radiative corrections to SM processes, etc.\,\cite{Harigaya:2015yaa}).

\vskip 0.5cm
\noindent
{\bf Acknowledgments}\\[0.1cm]
\noindent
This work is supported by the Grant-in-Aid for Scientific research from the Ministry of Education, Science, Sports, and Culture (MEXT), Japan No. 26104009 and 26287039 (S. M.), as well as by the World Premier International Research Center Initiative (WPI), MEXT, Japan (S. M., K. M. and Y. S. T.). The work of K.M. is supported in part by a JSPS Research Fellowships for Young Scientists. S. B acknowledges the support of the Indo French LIA THEP (Theoretical high Energy Physics) of the CNRS. S. B. acknowledges the hospitality of IPMU where the idea for this work was conceived, and also acknowledges the cluster facility at the Harish-Chandra Research Institute\,\url{http://www.hri.res.in/cluster/}. We also thank Daniel Schmeier for technical help regarding CheckMATE.

\newpage

\appendix

\section{Blind Spot Conditions}
\label{app: blind spots}

Explicit forms of the effective Yukawa and gauge couplings of the WIMP in Eqs\,(\ref{eq: y11}) and (\ref{eq: g11}) are derived in this appendix. These forms can in principle be obtained by calculating the mixing matrix $U_N$ explicitly. There is, however, a more efficient way to obtain the forms, and this is nothing but the derivation that we will introduce below.

\subsection{The Higgs blind spot condition}
\label{app: higgs blind spot}

Using the real orthogonal matrix $O_N$ instead of the could-be complex mixing matrix $U_N$, the definition of the effective Yukawa coupling of $N_i$ in Eq.\,(\ref{eq: def y11}) can be expressed as
\begin{align}
	y_{ii} = \frac{\eta_i^2}{\sqrt{2}} [-(O_N^T)_{i1} y_1 (O_N)_{2i} + (O_N^T)_{i1} y_2 (O_N)_{3i}].
	\label{app: def y11}
\end{align}
Since we assumed the CP invariance in the singlet-doublets WIMP model, the Yukawa coupling $y_{ii}$ is always real ($y_{ii} = y_{ii}^\ast$), so that all the neutral particles $N_i$s have their own scalar interactions ($h\,\bar{N}_i N_i$), while do not have the pseudoscalar ones ($ih\,\bar{N}_i \gamma_5 N_i$).

As a first step of the derivation, we define the $\epsilon$-modified mass matrix $M_N(\epsilon)$, which is obtained by extending the mass matrix $M_N$ in Eq.\,(\ref{eq: mass matrix}) by a small parameter $\epsilon$:
\begin{align}
	M_N (\epsilon) \equiv M_N + \delta M_N(\epsilon), \qquad
	\delta M_N(\epsilon) \equiv \begin{pmatrix} 0 & -y_1 \epsilon/\sqrt{2} & y_2 \epsilon/\sqrt{2} \\
	-y_1 \epsilon/\sqrt{2} & 0 & 0 \\ y_2 \epsilon/\sqrt{2} & 0 & 0 \end{pmatrix}.
	\label{app: MMM for y11}
\end{align}
The original mass matrix $M_N$ is obtained when $\epsilon = 0$, namely $\delta M_N(0) = 0$. Moreover, when $M_N$ is regarded as a function of $v$, $M_N(\epsilon)$ can be written as $M_N(\epsilon) = M_N|_{v \to v + \epsilon}$. Let us now think about the diagonalization of $M_N(\epsilon)$. Since $M_N(\epsilon)$ is a real symmetric matrix, it can be done by a real orthogonal matrix $O_N(\epsilon)$, where $O_N(\epsilon)$ is also expressed as $O_N(\epsilon) \equiv O_N + \delta O_N(\epsilon) = O_N|_{v \to v + \epsilon}$ with $\delta O_N(\epsilon)$ satisfying $\delta O_N(0) = 0$. The eigenvalue $\tilde{M}_{Ni}(\epsilon)$, which is obtained by diagonalizing $O_N(\epsilon)$, is again expressed as $\tilde{M}_{Ni}(\epsilon) \equiv \tilde{M}_{Ni} + \delta\tilde{M}_{Ni}(\epsilon) = \tilde{M}_{Ni}|_{v \to v + \epsilon}$ with $\tilde{M}_{Ni}$ being the eigenvalue of $M_N$, and therefore $\delta\tilde{M}_{Ni}(0) = 0$.

The next step is to expand the eigenvalue $\tilde{M}_{Ni}(\epsilon)$ by the small parameter $\epsilon$. With the use of the diagonalizing equation, $O_N^T(\epsilon)\,M_N(\epsilon)\,O_N(\epsilon) = {\rm diag}[\tilde{M}_{N1}(\epsilon), \tilde{M}_{N2}(\epsilon), \tilde{M}_{N3}(\epsilon)]$, the following relation can be derived in the $\epsilon$-dependent part of the equation:
\begin{align}
	\delta \tilde M_{Ni} (\epsilon) 
	= [ O_N^T\,\delta M_N(\epsilon)\,O_N ]_{ii}
	+\tilde{M}_{Ni} [ \delta O_N^T(\epsilon)\,O_N + O_N^T\,\delta O_N(\epsilon) ]_{ii}
	+ {\cal O}(\epsilon^2).
\end{align}
The orthogonal condition of the diagonalizing matrix, $O_N^T(\epsilon)\,O_N(\epsilon) = {\mathbb 1}$, leads to another equation $\delta O_N^T (\epsilon)\,O_N + O_N^T \,\delta O_N (\epsilon) = {\cal O}(\epsilon^2)$, so that the second term in the right hand side can be dropped. On the other hand, the first term in the same side turns out to have almost the same form of the effective Yukawa coupling $y_{11}$ defined in Eq.\,(\ref{app: def y11}). As a result, the effective coupling $y_{ii}$ can be expressed by using $\tilde{M}_{Ni} (\epsilon)$ as follows:
\begin{align}
	y_{ii}
	= \frac{\eta_i^2}{2}
	\left. \frac{\partial \delta \tilde{M}_{Ni}(\epsilon)}{\partial \epsilon} \right|_{\epsilon = 0}
	= \frac{\eta_i^2}{2}
	\left. \frac{\partial \tilde{M}_{Ni}(\epsilon)}{\partial \epsilon} \right|_{\epsilon = 0}
	= \frac{\eta_i^2}{2}
	\left. \frac{\partial \tilde{M}_{Ni}(\epsilon)}{\partial v} \right|_{\epsilon = 0}
	= \frac{\eta_i^2}{2}
	\frac{\partial \tilde{M}_{Ni}}{\partial v},
	\label{app: another form of y11}
\end{align}
where $\tilde{M}_{Ni}(\epsilon) = \tilde{M}_{Ni}|_{v \to v + \epsilon}$ has been used to change the variable from $\epsilon$ to $v$.

The last step is to compute the explicit form of $\tilde{M}_{Ni}$ (the derivative of $\tilde{M}_{Ni}(\epsilon)$ to be more precise) utilizing the eigenvalue equation, $\det\,[M_N - \tilde{M}_{Ni}\,\mathbb{1}] = 0$. Differentiating the eigenvalue equation with respect to $v$, we can obtain the following useful expression of $y_{ii}$ without computing the mixing matrix $U_N$ explicitly:
\begin{align}
	y_{ii} =
	- \eta_i^2 \frac{y^2\,v\,(M_D \sin 2\theta + \tilde{M}_{Ni})}
	{2M_D^2 + y^2 v^2 + 4 M_S \tilde{M}_{Ni} - 6 \tilde{M}_{Ni}^2}.
	\label{app: y11}
\end{align}
Because $M_{Ni} = \eta_i^2\,\tilde{M}_{Ni}$, it is nothing but the one in Eq.\,(\ref{eq: y11}) by taking $i = 1$.

Here, we also derive another fact `{\boldmath $\tilde{M}_{N1} = M_S$}' when the Higgs blind spot condition holds ($y_{11} = 0$). From the expression of $y_{11}$ in Eq.\,(\ref{app: another form of y11}), the blind spot condition is equivalent to $\partial \tilde{M}_{N1}/\partial v = 0$ for an arbitrary $v$, meaning that $\tilde{M}_{N1}$ is constant with respect to $v$ whenever the condition holds. Let us now take $v = 0$. Then, $M_N(0)$ has a simple form and its eigenvalues are obtained as $M_S$, $M_D$ and $-M_D$, where those must be the eigenvalues of $M_N$ with $v$ being 246\,GeV, namely $\tilde{M}_{Ni} = M_S$, $M_D$ or $-M_D$. When $M_S \leqslant |M_D|$, the mass of the lightest neutral particle $N_1$ becomes {\boldmath $M_{N1} = \eta_1^2 \tilde{M}_{N1} = \eta_1^2 M_S$}, so that {\boldmath $\eta_1 = 1$}. When $|M_D| \leqslant M_S$, we have two choices, $\tilde{M}_{N1} = M_D$ or $-M_D$. However, the first one is forbidden because it cannot satisfy the condition, $M_D \sin 2 \theta + \tilde M_{Ni} = 0$, due to $\sin 2\theta \geqslant 0$. On the other hand, the second one can satisfy the condition when $\sin 2\theta = 1$, namely $\theta = \pi/4$. However, $\theta = \pi/4$ turns out to result in an SU(2)$_L$ doublet WIMP ($R_S = 0$), so that the blind spot with $\tan \theta =1$ is out of our interest, say the WIMP in the well-tempered regime\,(\ref{eq: well-tempered condition}).

\subsection{The Z boson blind spot condition}
\label{app: z blind spot}

As in the case of the effective Yukawa coupling $y_{ii}$, the effective gauge coupling of $N_i$ in Eq.\,(\ref{eq: def g11}) can also be expressed using the real orthogonal matrix $O_N$:
\begin{align}
	g_{ii} = \frac{g}{4 c_W} [(O_N^T)_{i2} (O_N)_{2i} - (O_N^T)_{i3} (O_N)_{3i}].
	\label{app: def g11}
\end{align}
The coupling $g_{ii}$ becomes always real ($g_{ii} = g_{ii}^\ast$), so that all the neutral particles $N_i$s have their own axial vector current interactions ($\bar{N}_i \slashed{Z} \gamma_5 N_i$), while never have the vector current ones ($\bar{N}_i \slashed{Z} N_i$), as expected from the Majorana nature of $N_i$s.

The explicit form of the effective gauge coupling can be obtained by exactly the same method as the one for the effective Yukawa coupling. Instead of the modified mass matrix shown in Eq.\,(\ref{app: MMM for y11}), we define the following $\epsilon$-modified mass matrix this time:
\begin{align}
	M_N (\epsilon) \equiv M_N + \delta M_N(\epsilon), \qquad
	\delta M_N(\epsilon) \equiv {\rm diag}(0, \epsilon, -\epsilon).
	\label{app: MMM for g11}
\end{align}
We first diagonalize this modified mass matrix, and expand its eigenvalue $\tilde{M}_{Ni}(\epsilon)$ in terms of the small parameter $\epsilon$. We then obtain the relation, $\delta \tilde{M}_{Ni} (\epsilon) = [O_N^T\,\delta \tilde{M}_{N} (\epsilon)\,O_N ]_{ii} + {\cal O} (\epsilon^2)$. The first term in the right hand side has almost the same form of $g_{ii}$ in Eq.\,(\ref{app: def g11}), so that we have an expression of $g_{ii}$, which is similar to the one for $y_{ii}$ in Eq.\,(\ref{app: another form of y11}):
\begin{align}
	g_{ii} &= \frac{g}{4 c_W} 
	\left. \frac{\partial \tilde{M}_{Ni}}{\partial \epsilon} \right|_{\epsilon = 0}.
\end{align}
The derivative of $\tilde{M}_{Ni}$ is obtained by the eigenvalue equation, $\text{det}\,[M_N(\epsilon) - \tilde{M}_{Ni}(\epsilon)\,\mathbb{1}] = 0$. Differentiating the equation with respect to $\epsilon$, we get the following formula:
\begin{align}
	g_{ii} = - \frac{y^2\,v\,m_Z\,\cos 2\theta}
	{2\,(2M_D^2 + y^2 v^2 + 4 M_S \tilde{M}_{Ni} - 6 \tilde{M}_{Ni}^2}.
\end{align}
Remembering $M_{Ni} = \eta_i^2\,\tilde{M}_{Ni}$ again, it is just the one in Eq.\,(\ref{eq: g11}) by taking $i = 1$.

\section{14\,TeV LHC studies}
\label{app: lhc14}

For the search on the singlet-doublets WIMP model at the 14\,TeV LHC with 3\,ab$^{-1}$ integrated luminosity, namely the high-luminosity run of the LHC (HL-LHC), we implement all the cuts listed in the ATLAS prediction study for the same final state. We generate the events using {\tt MadGraph5}\,\cite{Alwall:2014hca} and shower them by {\tt Pythia6}\,\cite{Sjostrand:2006za}. The detector level analysis is performed using {\tt Delphes3}\,\cite{deFavereau:2013fsa} with the detector cards tuned to mimic the simulation of Ref.\,\cite{ATL-PHYS-PUB-2014-010}. The final state is a same flavor opposite sign (SFOS) pair of leptons ($e^+e^-$ or $\mu^+\mu^-$) and missing transverse energy ($\slashed{E}_T$). Following are the cuts used in our study:
\begin{enumerate}
	\setlength{\itemsep}{0cm}
	\item Trigger cuts on leptons: $p_T(\ell) > 10$\,GeV and $|\eta| < 2.47\,(2.4) $ for $e\,(\mu)$.
 	\item Jets reconstruction: Anti-$k_T$\,\cite{Cacciari:2008gp} with FastJet\,\cite{Cacciari:2011ma} with $p_T > 20$\,GeV and $\eta < 2.5$.
 	\item Jets tagged as $b$-tagged jets have an average efficiency of 70\% and a light flavour jet misidentification probability is 1\%.
 	\item Leptons with $m_{\rm SFOS} < 12$\,GeV are removed to reduce low resonance backgrounds
 	\item Lepton isolation: Demanding the scalar sum of tracks with $p_T > 1$\,GeV should not be more than 15\% of the sum within a cone of radius $\Delta R = 0.4$ around the lepton. 
 	\item Leptons with $\Delta R < 0.1$ are discarded.
 	\item Events where a lepton comes closer to a jet than $\Delta R < 0.4$ are discarded.
 	\item Events are selected with exactly 3 leptons and an SFOS pair.
 	\item Events with $b$-tagged jets are discarded to reduce $t\bar{t}$ and $t\bar{t}+V$ backgrounds.
 	\item At least a SFOS pair is required to peak around $m_Z$, $i.e.$ $|m_{\rm SFOS} - m_Z| < 10$\,GeV.
 	\item $p_T(\ell_1, \ell_2, \ell_3) > 50$\,GeV is also applied.
 	\item Lastly, large $\slashed{E}_T$ and $m_T$ requirements are made to define four signal regions (SR), $viz.$
 	SR A: $\slashed{E}_T > 250$\,GeV and $m_T > 150$\,GeV,
 	SR B: $\slashed{E}_T > 300$\,GeV and $m_T > 200$\,GeV,
 	SR C: $\slashed{E}_T > 400$\,GeV and $m_T > 200$\,GeV,
 	SR D: $\slashed{E}_T > 500$\,GeV and $m_T > 200$\,GeV.
\end{enumerate}
Even though here the cross-sections are larger compared to the 8 TeV analysis, the signal regions are inadequate to put strong constraints on the WIMP model. One possibly needs to tune the signal regions to suit our needs. We must mention that for the entire parameter space allowed after passing through XENON1T expected limits, the partonic cross-sections for the aforementioned final state with loose trigger cuts vary from $\sim 0.002$\,fb to a maximum of $\sim 0.1$\,fb. However, on implementing the cuts we are left with tiny $S/B$ and $S/\sqrt{S+B}$. Hence, we are unable to rule out any part of the parameter space even with the HL-LHC.

\section{Supplementary figures}
\label{app: figures}

For the sake of completeness, we provide several figures in this appendix which we obtain by numerically scanning the parameter space of the singlet-doublets WIMP model. These figures play a supplementary role for the discussion in the main text. Below, three sets of the figures are shown which explain the present status, the near future prospects and the future prospects of the WIMP model, respectively. Six figures are depicted in each set, where the profile likelihood contours on the planes of all possible combinations of the input parameters, $M_S$, $M_D$, $y$ and $\cot\theta$, are shown.

\begin{figure}[H]
	\centering
	\begin{tabular}{ccc}
	\subfloat[\label{appfig: MS_MD_P}]{\includegraphics[width=0.3\textwidth]{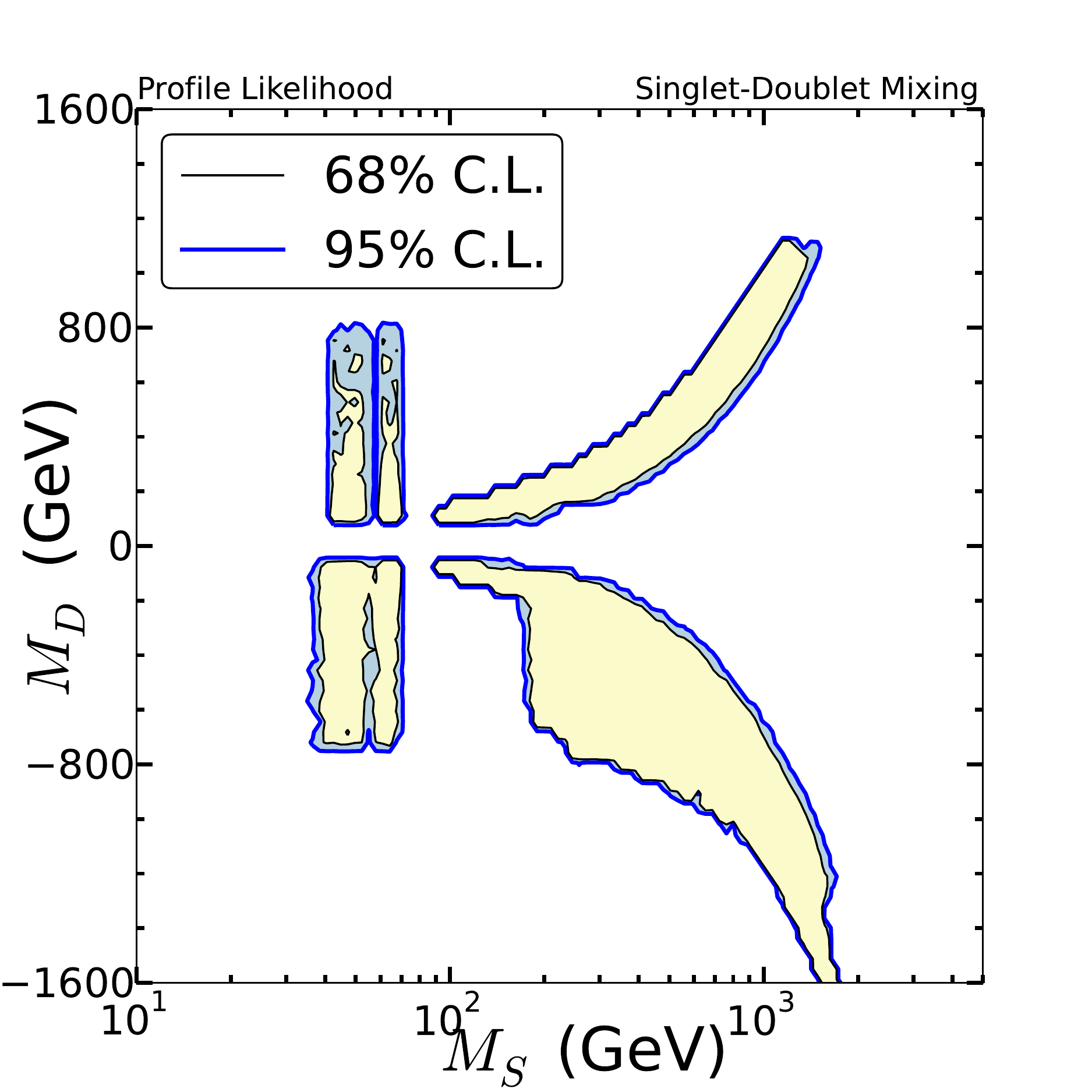}} & & \\
	\subfloat[\label{appfig: MS_Y_P}]{\includegraphics[width=0.3\textwidth]{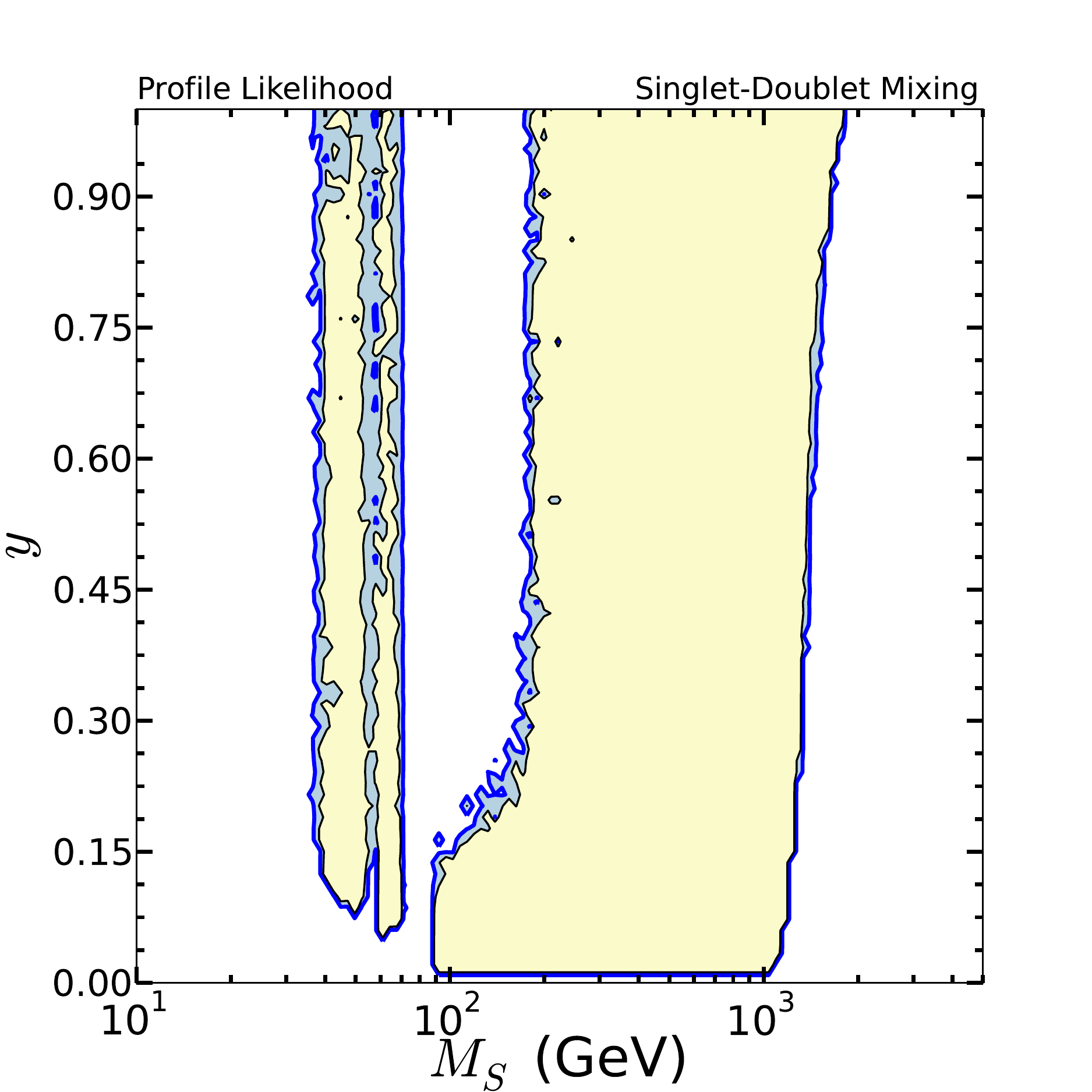}} &
	\subfloat[\label{appfig: MD_Y_P}]{\includegraphics[width=0.3\textwidth]{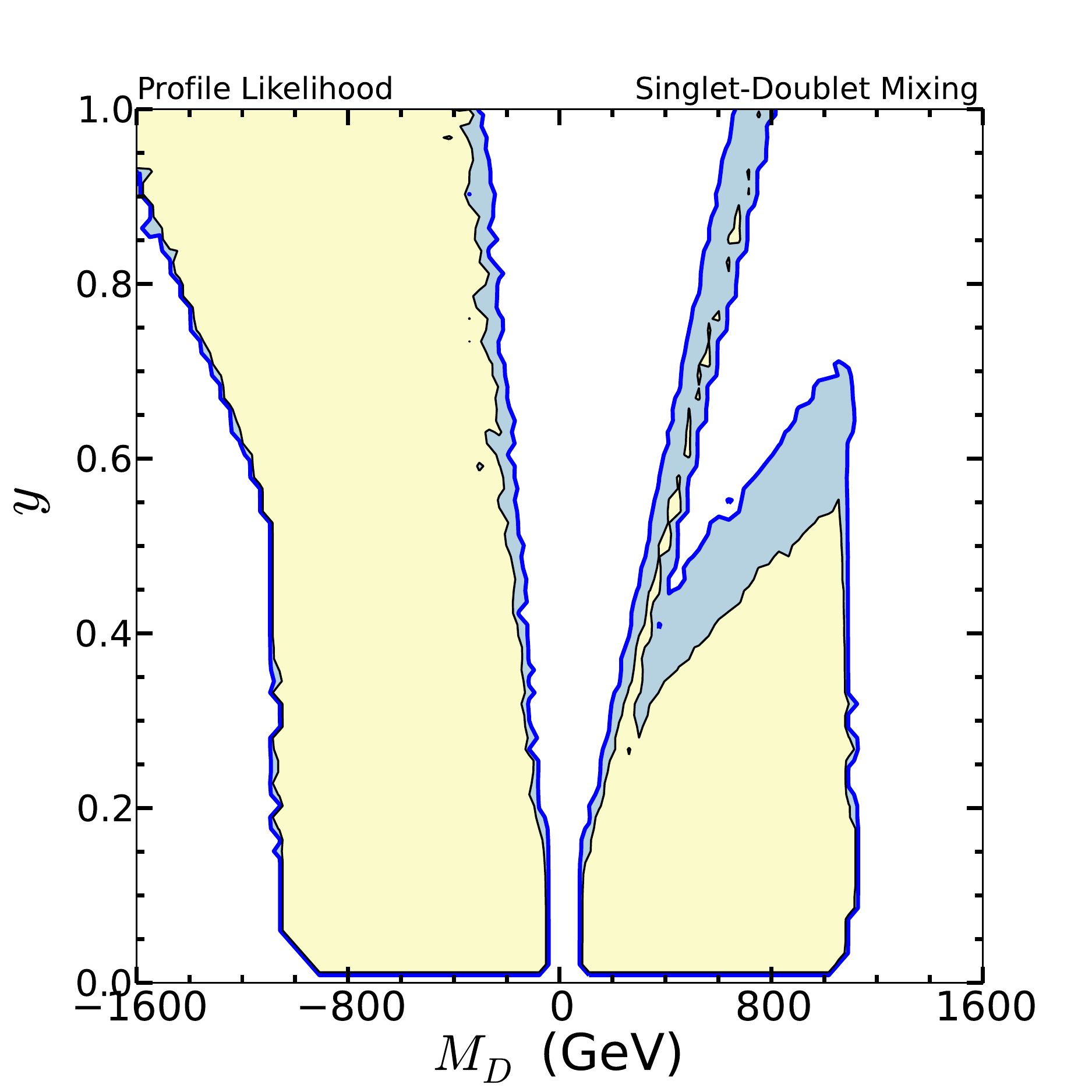}} & \\
	\subfloat[\label{appfig: MS_COT_P}]{\includegraphics[width=0.3\textwidth]{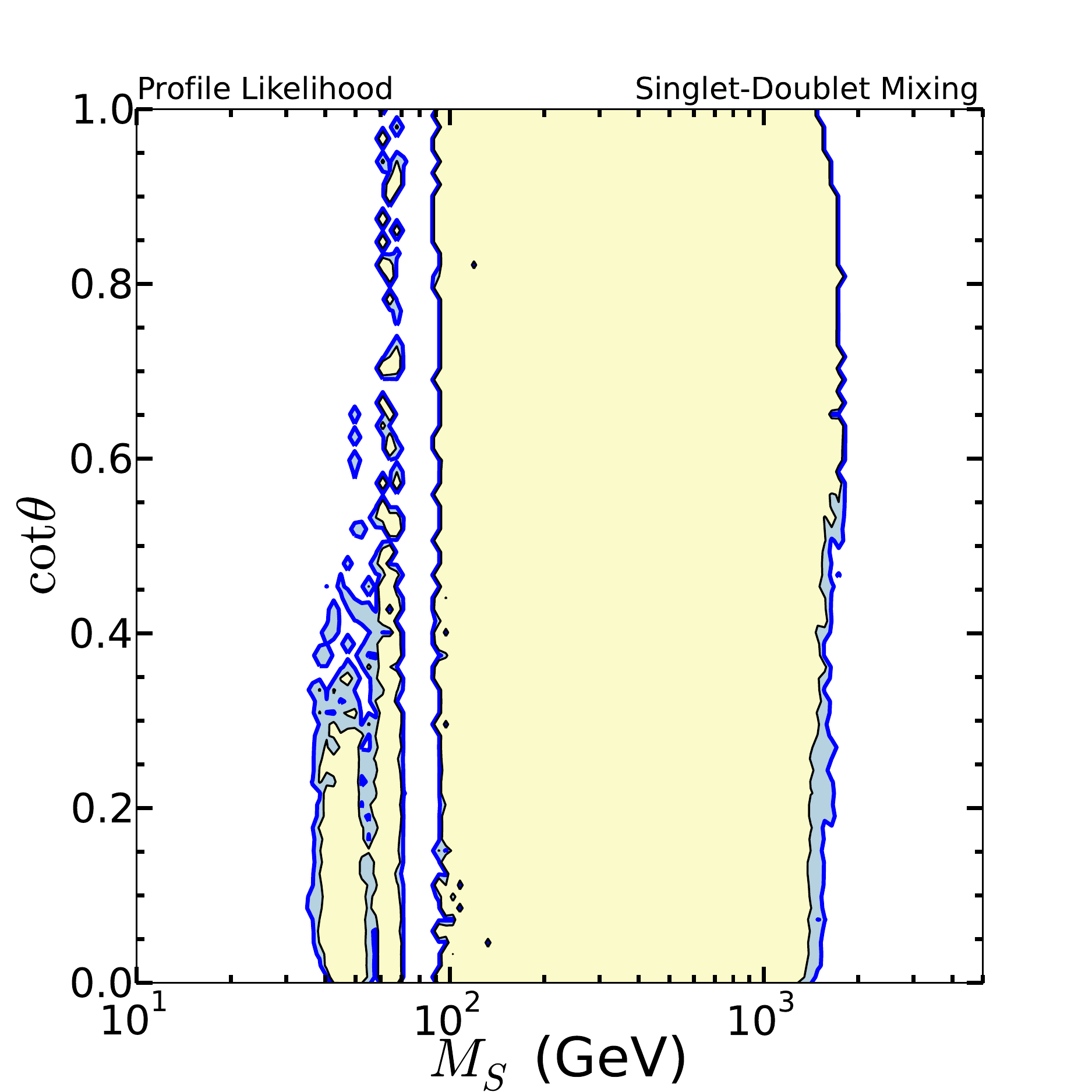}} & 
	\subfloat[\label{appfig: MD_COT_P}]{\includegraphics[width=0.3\textwidth]{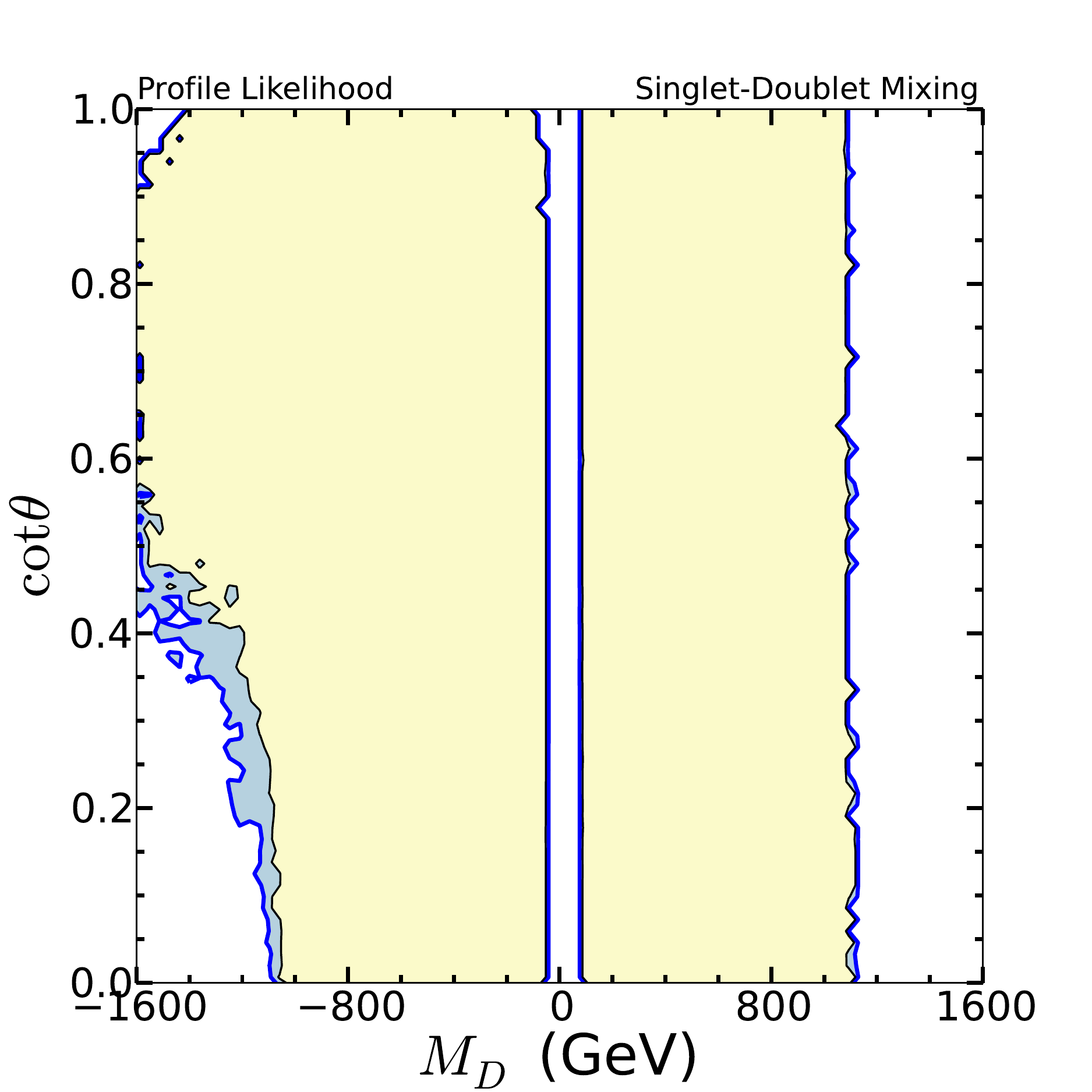}} &  
	\subfloat[\label{appfig: Y_COT_P}]{\includegraphics[width=0.3\textwidth]{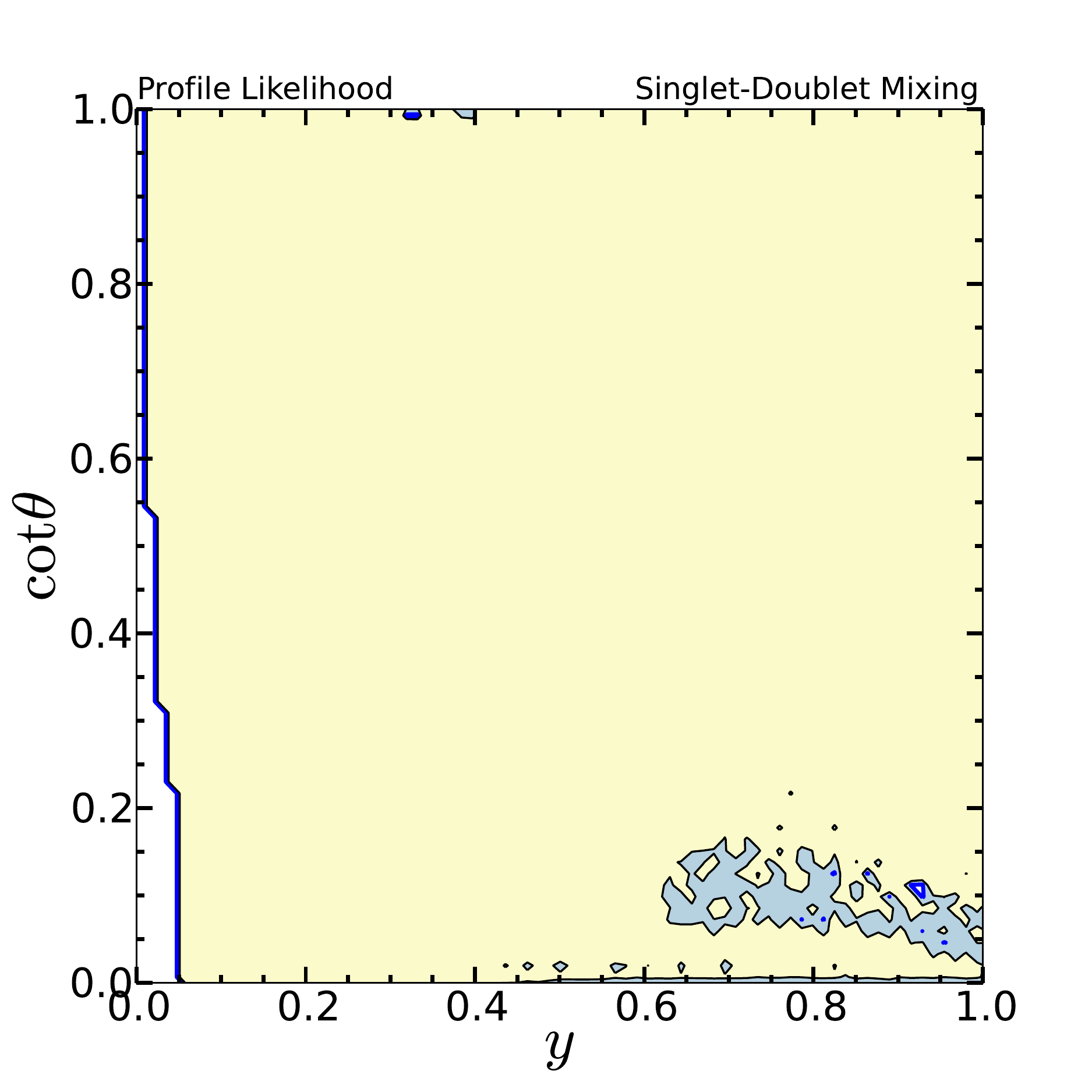}} 
	\end{tabular}
	\caption{\sl \small Profile likelihood contours of our input parameter space at present.}
	\label{appfig: current}
\end{figure}

\begin{figure}[H]
	\centering
	\begin{tabular}{ccc}
	\subfloat[\label{appfig: MS_MD_NF}]{\includegraphics[width=0.3\textwidth]{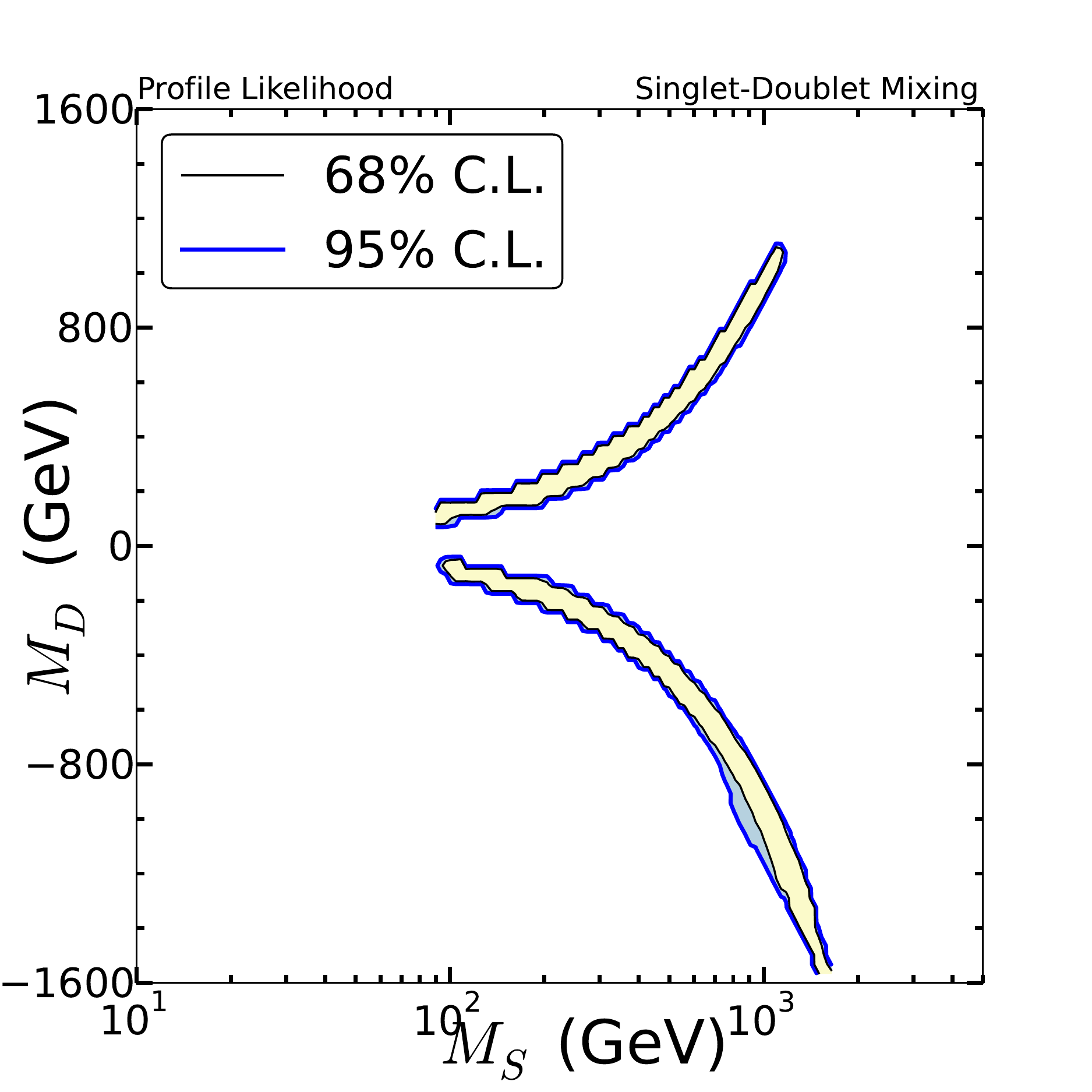}} & & \\
	\subfloat[\label{appfig: MS_Y_NF}]{\includegraphics[width=0.3\textwidth]{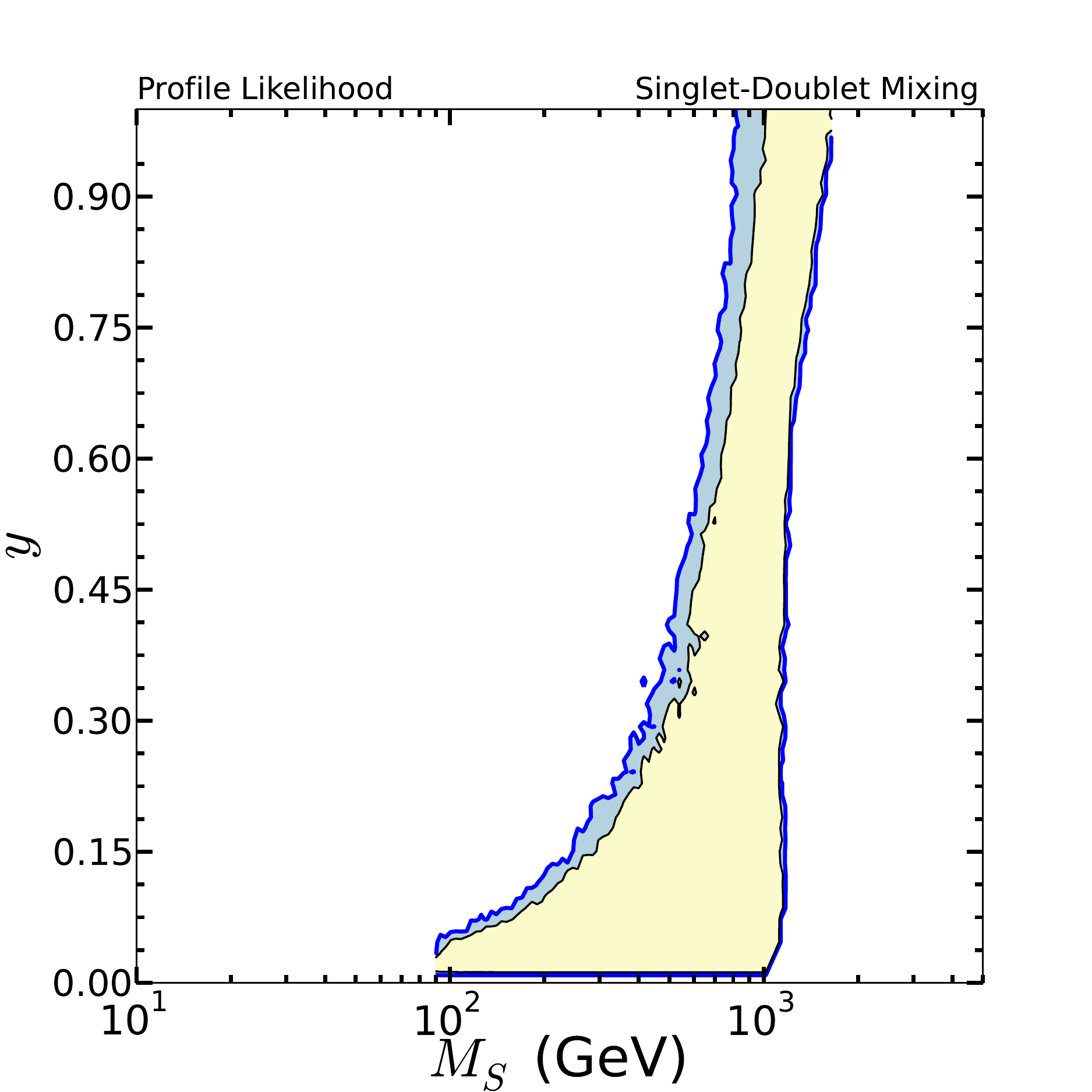}} &
	\subfloat[\label{appfig: MD_Y_NF}]{\includegraphics[width=0.3\textwidth]{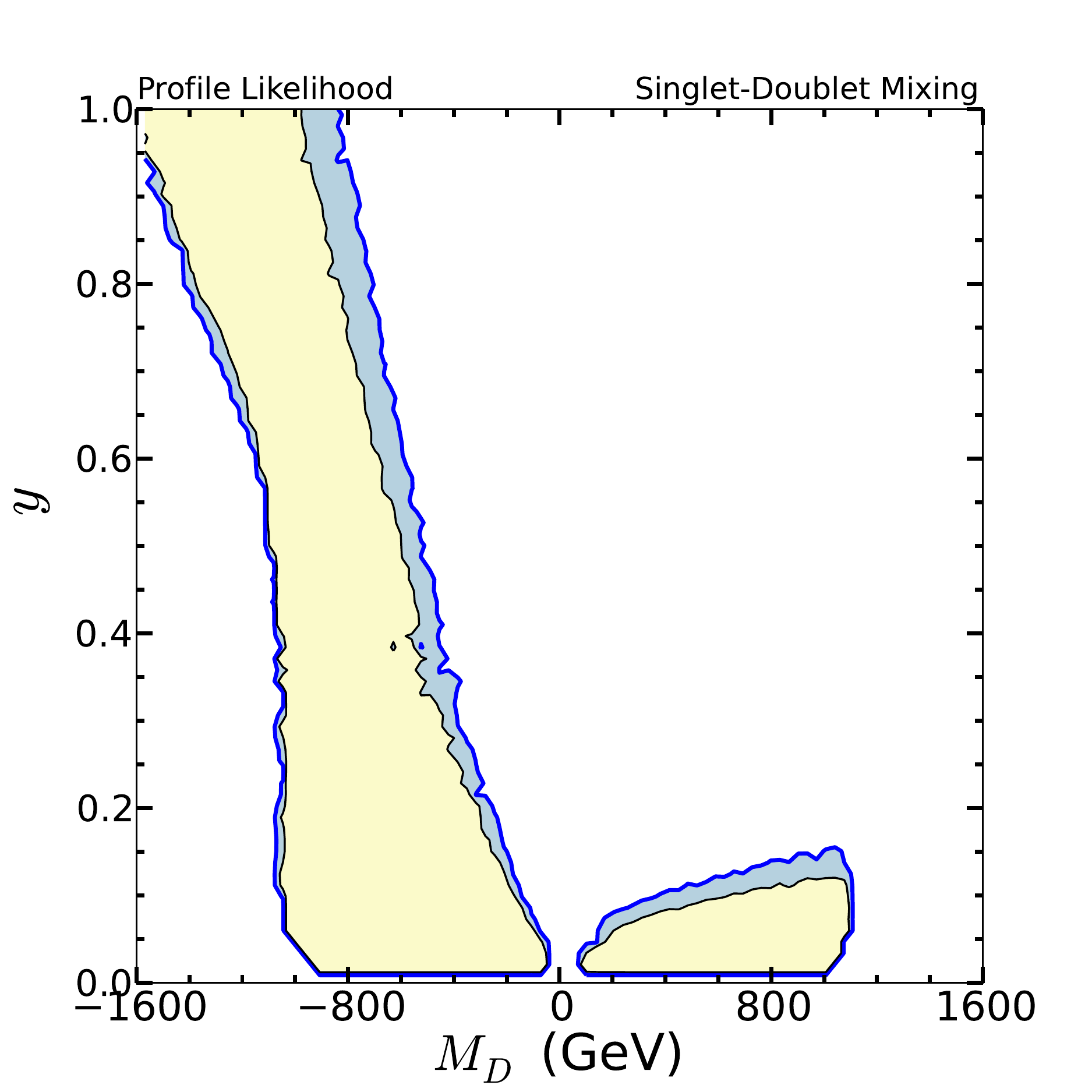}} & \\
	\subfloat[\label{appfig: MS_COT_NF}]{\includegraphics[width=0.3\textwidth]{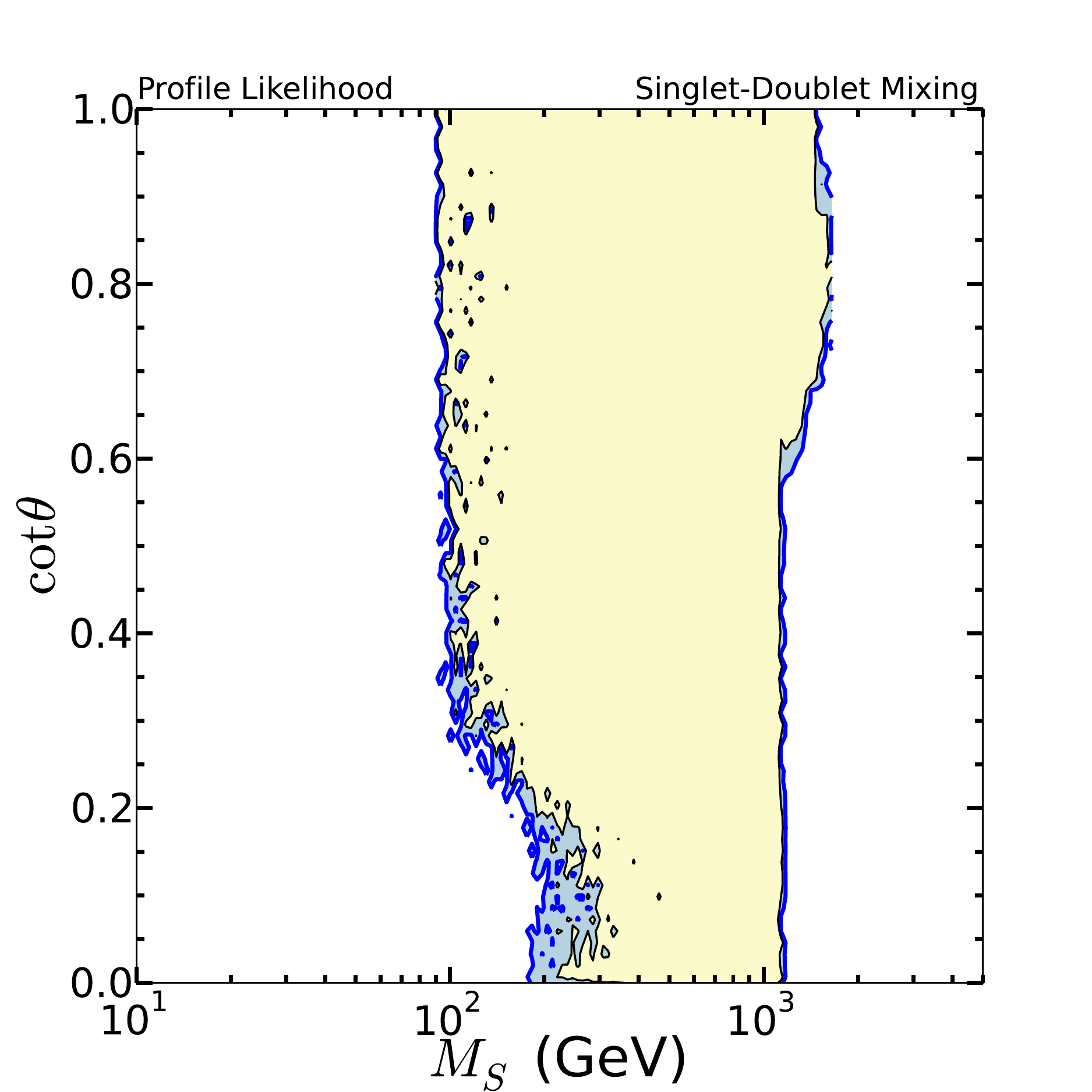}} & 
	\subfloat[\label{appfig: MD_COT_NF}]{\includegraphics[width=0.3\textwidth]{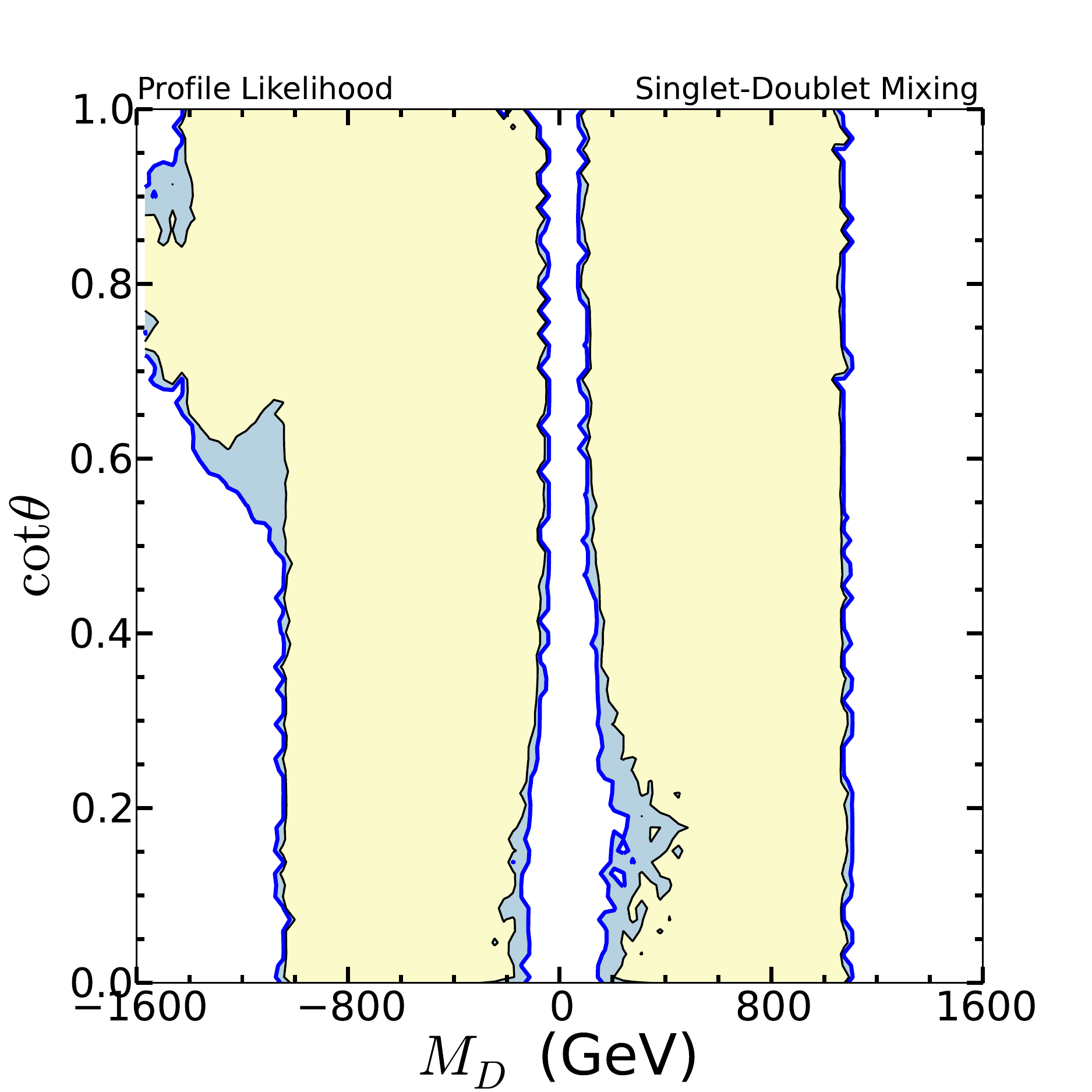}} &  
	\subfloat[\label{appfig: Y_COT_NF}]{\includegraphics[width=0.3\textwidth]{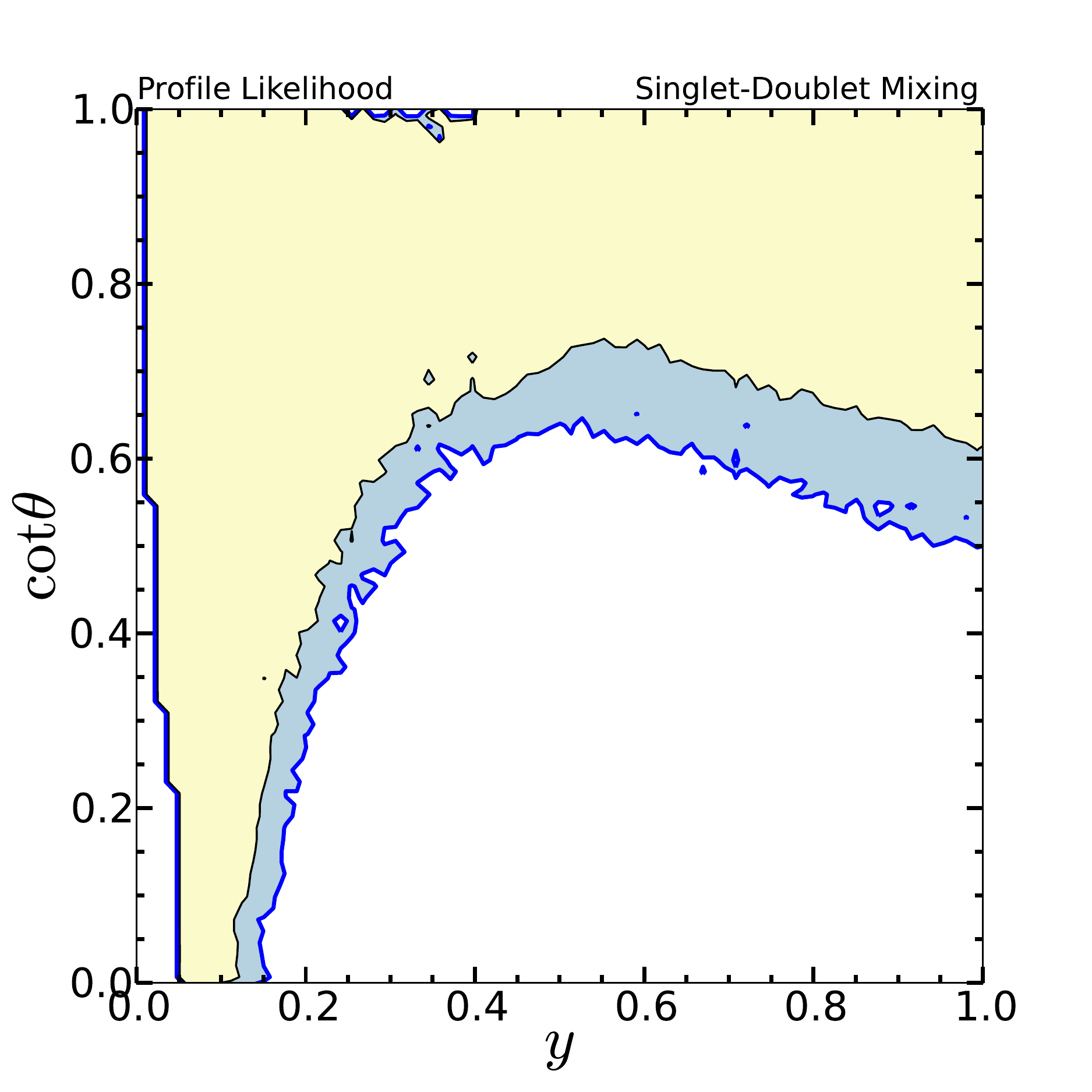}} 
	\end{tabular}
	\caption{\sl \small Profile likelihood contours of our input parameter space in the near future.}
	\label{appfig: near future}
\end{figure}

\begin{figure}[H]
	\centering
	\begin{tabular}{ccc}
	\subfloat[\label{appfig: MS_MD_F}]{\includegraphics[width=0.3\textwidth]{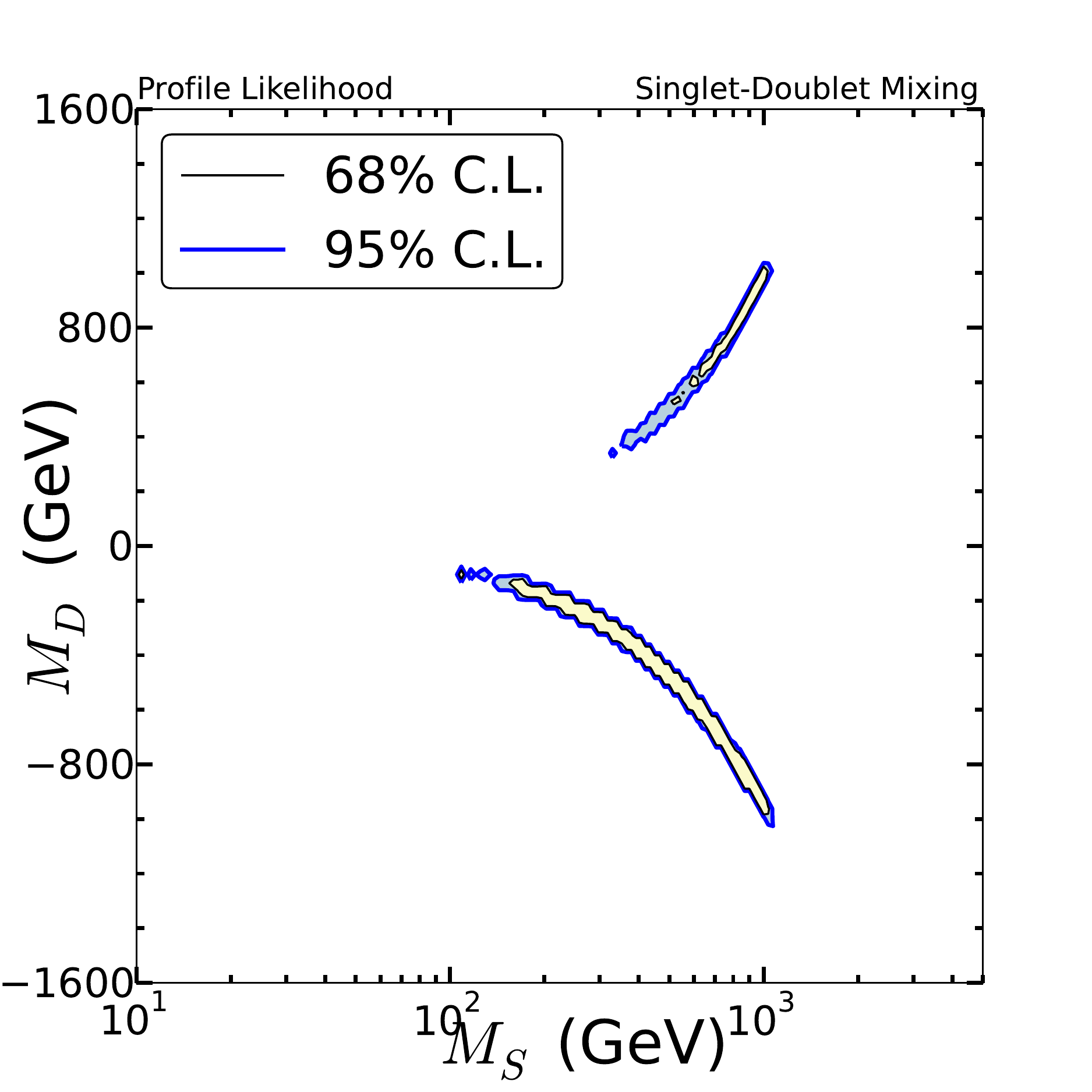}} & & \\
	\subfloat[\label{appfig: MS_Y_F}]{\includegraphics[width=0.3\textwidth]{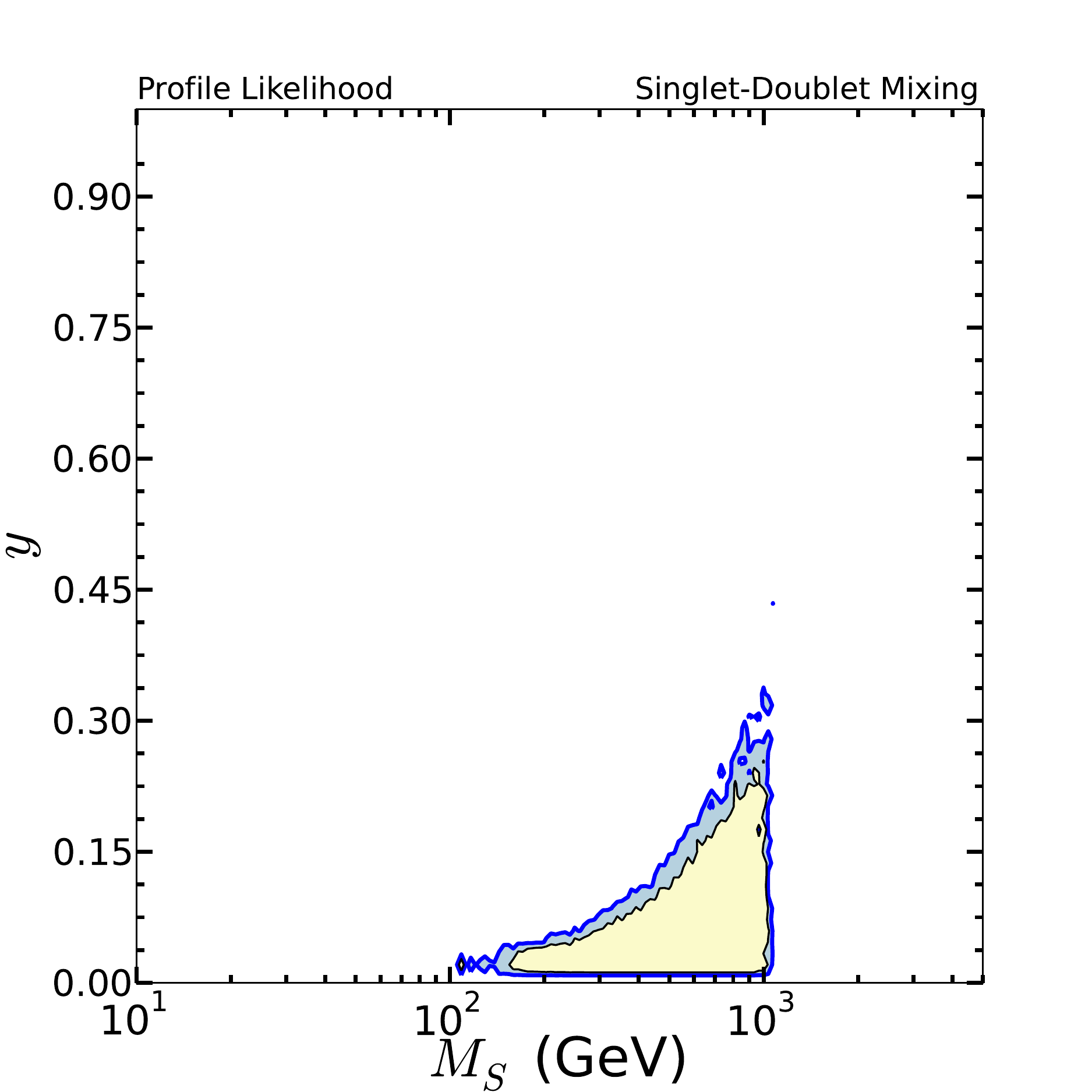}} &
	\subfloat[\label{appfig: MD_Y_F}]{\includegraphics[width=0.3\textwidth]{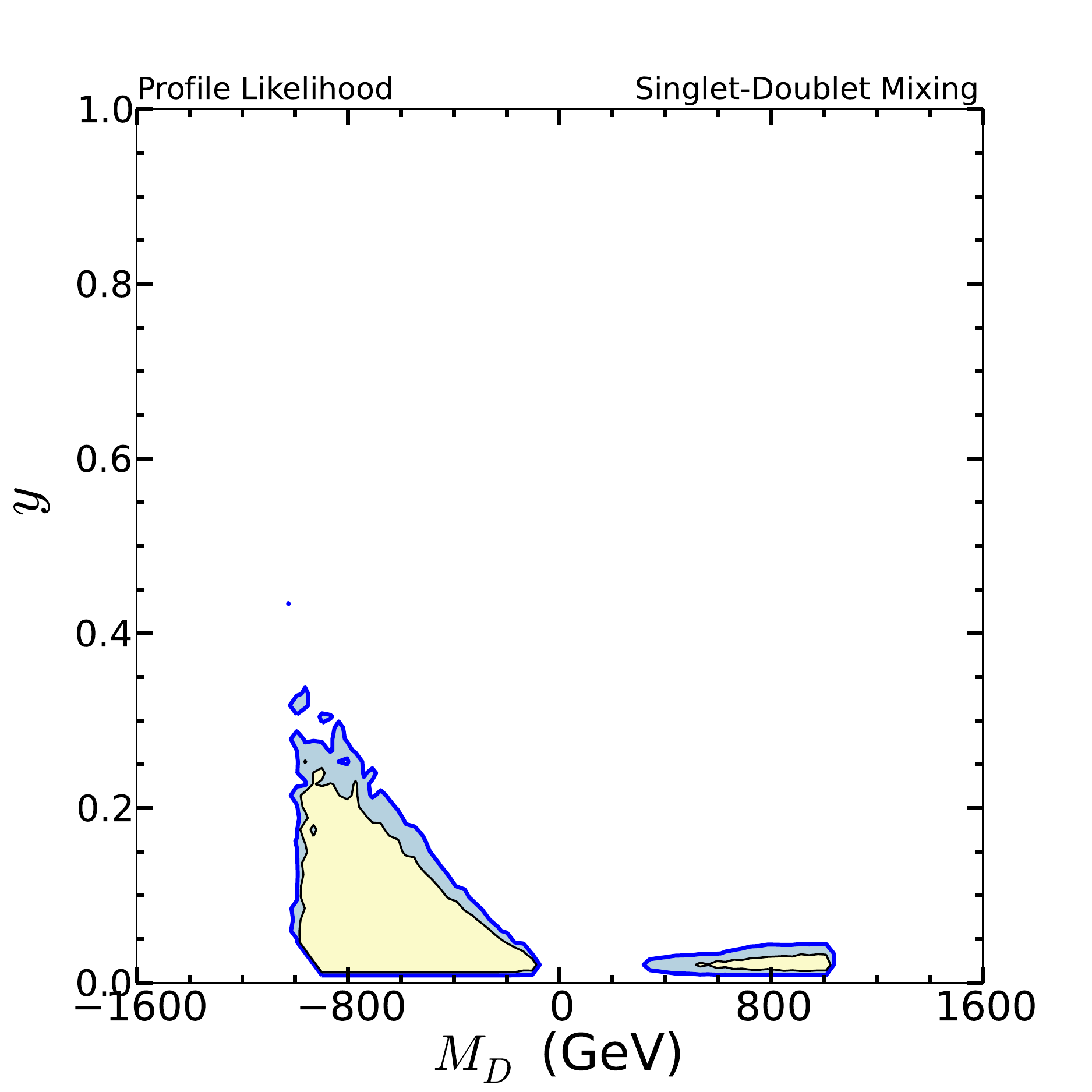}} & \\
	\subfloat[\label{appfig: MS_COT_F}]{\includegraphics[width=0.3\textwidth]{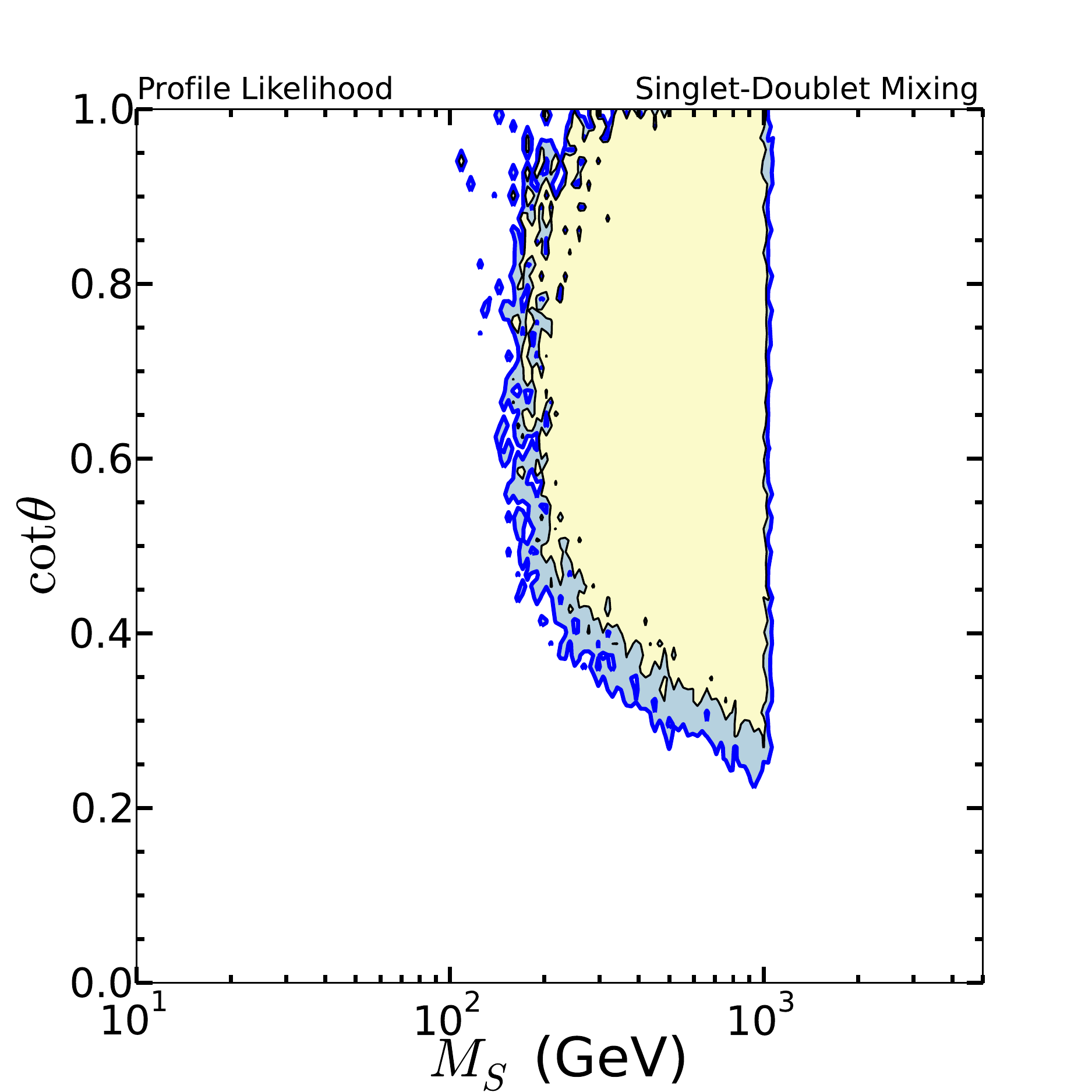}} & 
	\subfloat[\label{appfig: MD_COT_F}]{\includegraphics[width=0.3\textwidth]{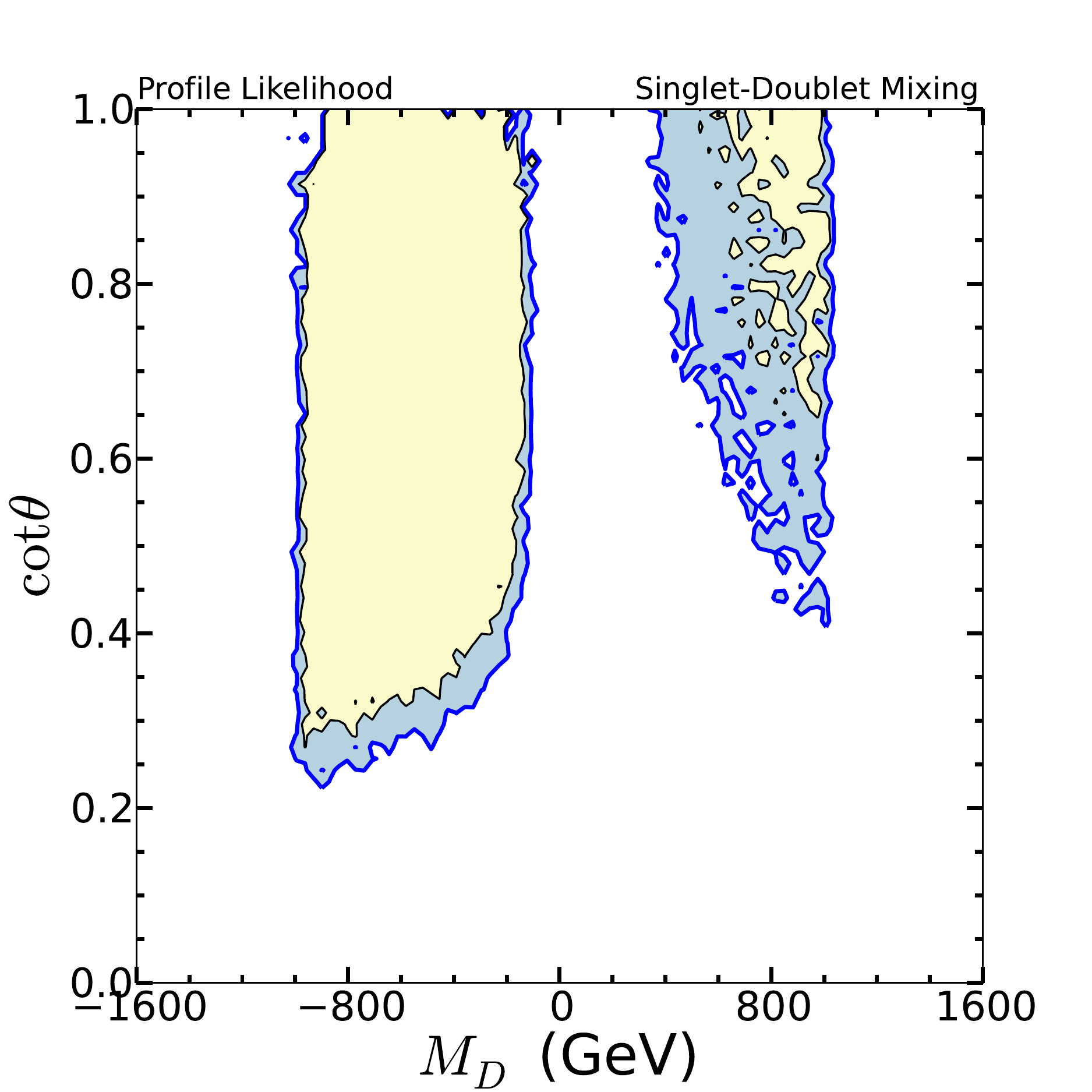}} &  
	\subfloat[\label{appfig: Y_COT_F}]{\includegraphics[width=0.3\textwidth]{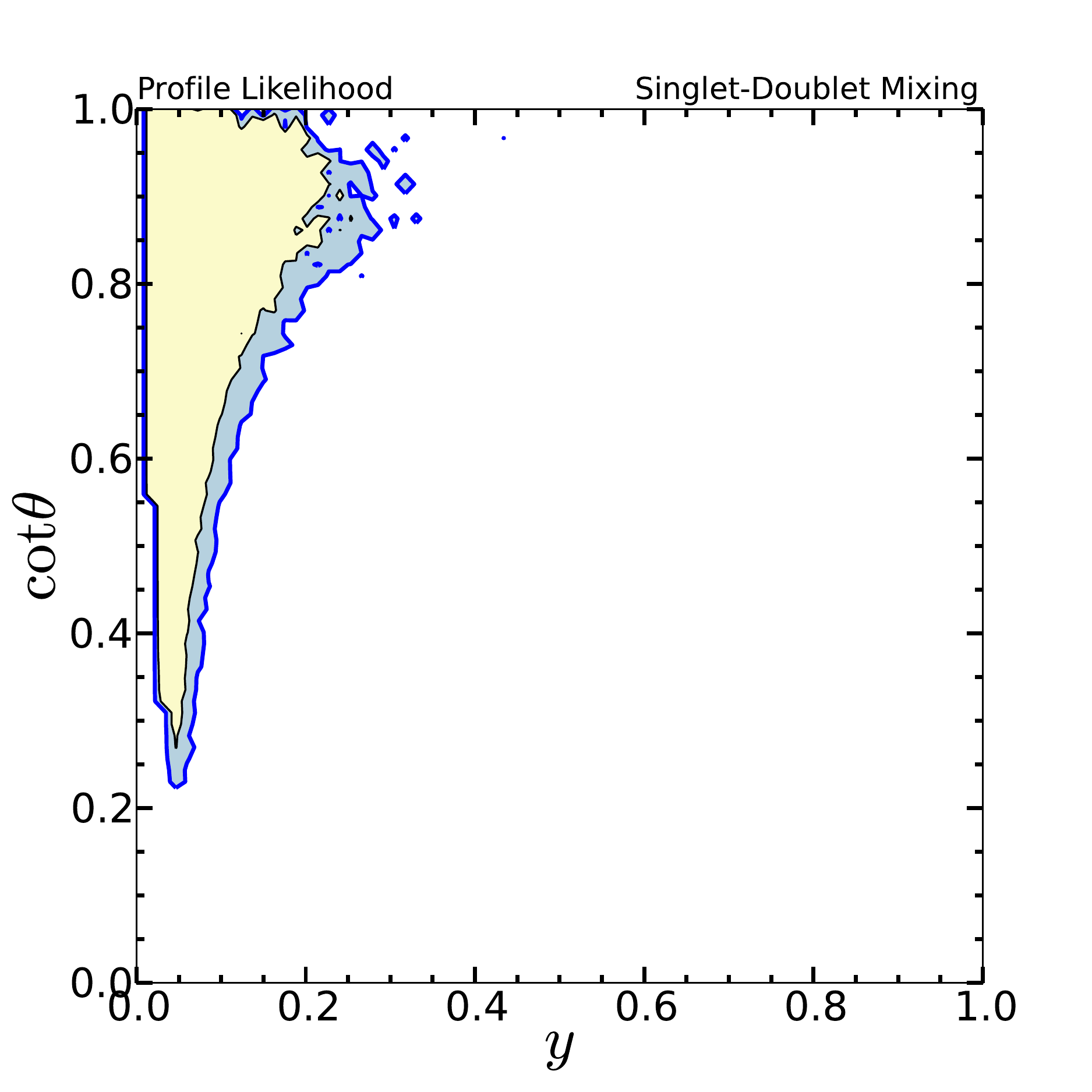}} 
	\end{tabular}
	\caption{\sl \small Profile likelihood contours of our input parameter space in the future.}
	\label{appfig: future}
\end{figure}

\small
\bibliographystyle{utphys}
\bibliography{refs}
\end{document}